# Moving beyond the classic difference-in-differences model: A simulation study comparing statistical methods for estimating effectiveness of state-level policies

Beth Ann Griffin
RAND Corporation, Arlington, VA 22202

Megan S. Schuler
RAND Corporation, Boston, MA 02116

Elizabeth A. Stuart
Johns Hopkins Bloomberg School of Public Health, Baltimore, MD 21205

Stephen Patrick
Vanderbilt University Medical Center and School of Medicine, Nashville, TN 37232

Elizabeth McNeer
Vanderbilt University Medical Center, Nashville, TN 37232

Rosanna Smart
RAND Corporation, Santa Monica, CA 90401

David Powell
RAND Corporation, Arlington, VA 22202

Bradley D. Stein
RAND Corporation, Pittsburgh, PA 15213

Terry Schell
RAND Corporation, Santa Monica, CA 90401

Rosalie Liccardo Pacula
University of Southern California, Los Angeles, CA 90089

**Corresponding Author:**
Beth Ann Griffin
1200 South Hayes Street
Arlington, VA 22202
bethg@rand.org
703.413.1100, x5188

**Abstract**

**Background:** Evaluations of state-level policies are central for identifying effective policies and informing policymakers' decisions. State-level policy evaluations commonly employ a difference-in-differences (DID) study design; yet within this framework, statistical model specification varies notably across studies. More guidance is needed about which set of statistical models perform best when estimating the impacts of state-level policies on outcomes.

**Methods:** Motivated by applied state-level opioid policy evaluations, we implemented an extensive simulation study to compare the statistical performance of multiple variations of the two-way fixed effect models traditionally used for DID under a range of simulation conditions. We also explored the performance of autoregressive (AR) and GEE models. We simulated policy effects onto annual state-level opioid mortality rates and assessed statistical performance using various metrics, including directional bias, magnitude bias, and root mean squared error. We also report Type I error rates and the rate of correctly rejecting the null hypothesis (e.g., power), given the prevalence of frequentist null hypothesis significance testing in the applied literature.

**Results:** While most linear models resulted in minimal bias, non-linear models and population-weighted versions of classic linear two-way fixed effect and linear GEE models yielded considerable bias (60 to 160%). Further, root mean square error was minimized by linear AR models when examining crude mortality rates and by negative binomial models when examining raw death counts. In the context of frequentist hypothesis testing, many models yielded high Type I error rates and very low rates of correctly rejecting the null hypothesis (< 10%), raising concerns of spurious conclusions about policy effectiveness in the opioid literature. When considering performance across models, the linear AR models were optimal in terms of directional bias, root mean squared error, Type I error, and correct rejection rates.

**Conclusions:** The findings highlight notable limitations of commonly used statistical models for DID designs, designs widely used in opioid policy studies and in state policy evaluations more broadly. In contrast, the optimal model identified (the AR model) is rarely utilized in state-policy evaluation. We urge applied researchers to move beyond the classic DID paradigm and adopt the use of AR models.

# 1. BACKGROUND

Evaluations of state-level policies are central to identifying effective policies and informing policymakers' decisions, yet the methodological rigor of published studies varies (see Schuler, Heins (1) for a review of the opioid policy literature). State-level policy evaluations commonly employ a difference-in-differences (DID) study design; yet within this framework, statistical model specification varies notably across studies. The choice of model specification as well as other factors – including low outcome occurrence rates (e.g., opioid mortality), sample size (both the number of policy states as well as the number of time points available), and differences across states prior to policy adoption – can impact the accuracy and precision of effect estimates. Although numerous publications provide analytic guidance for policy evaluations using longitudinal data (2-6) methodological best practices have not been fully adopted by applied researchers. Furthermore, there have been no comprehensive examinations of the relative performance of commonly used statistical models under conditions that mimic those encountered in actual state policy evaluation settings.

A DID study design, broadly defined, has become dominant in the health care policy literature when using longitudinal data to evaluate the impact of state-level policies (7, 8). A DID design compares the outcomes observed among a group exposed to the policy of interest (treatment group) and an unexposed comparison group both across timepoints prior to policy implementation (first difference) and after policy implementation (second difference) – the policy effect is estimated as the difference between the first and second differences, hence "difference-in-differences" (7). However, a growing number of studies highlight challenges and limitations of a DID design, particularly when the key DID assumptions do not hold (4, 7, 9-11) or when sample size is limited (12). Additionally, it has been well-established that standard error corrections that adjust for violations of the assumed independence of the repeated measures in longitudinal datasets are needed to obtain accurate Type I error rates (13-17). Despite the wealth of knowledge concerning challenges of and best practices for DID designs in various settings, the applied literature largely does not reflect these insights (18-21).

With the aim of promoting adoption of more robust statistical methods in health policy research, the present study empirically compares the performance of multiple variations of the two-way fixed effect model traditionally used in the context of a DID design for state-level policy evaluation. Our motivating context is the ongoing U.S. opioid crisis, which claimed over 50,000 lives in 2019 alone (22) and has spurred states to adopt a myriad of opioid-related policies and initiatives. The urgency of the opioid crisis necessitates that accurate, robust statistical methods are utilized to identify effective state policies, yet our recent review of the "state of the science" of the opioid-policy literature highlighted that methodological rigor varied notably across studies (1). Applied researchers would benefit from additional, accessible guidance regarding the multitude of analytic choices both in the context of opioid-policy evaluations and state-level policy evaluations more generally. We are aware of only one other study considering relative performance across statistical methods in the context of health policy – that study compared analytic approaches for evaluating state gun policy laws on gun-related mortality, another high-stakes health policy setting (19). While in some ways the settings are similar in terms of longitudinal state-level outcomes, the conclusions may differ due to differences in the underlying outcome distributions (e.g., opioid related mortality is more highly skewed outcome than total firearm deaths).

The present study seeks to provide needed guidance about which set of statistical models commonly used in evaluations of state-level opioid policies with a DID study design perform best when estimating the impacts of state-level opioid policies on opioid-related mortality, with lessons that most likely apply to state policy evaluations more broadly. Using a simulation study based on observed state-level opioid mortality, we assessed statistical performance using various metrics, including directional bias, magnitude bias, and root mean squared error; we additionally report Type I error and the rate of correctly rejecting the null hypothesis, given the prevalence of frequentist null hypothesis significance testing (NHST) in the applied literature. Our findings indicate that some commonly-used methods have poor statistical performance, which has implications for interpreting the existing literature as well as conducting rigorous future evaluation studies. Our discussion provides important insights to statisticians and researchers regarding methods to estimate policy effects, and highlights that there is still methodological development needed to address the challenges of rigorous policy effect estimation in the context of complex policy settings.

## 2. METHODS

The data structure, simulation conditions, empirical models considered in our simulation study are detailed below.

### 2.1 Data Structure

The data structure we considered in this study was longitudinal, repeated annualized measures at the state level. The outcome considered was opioid-related mortality, measured annually in each state over 18 years, providing 50*18=900 total observations, clustered within states. We did not consider the existence of individual-level data within the aggregate state level data.

### 2.2 Empirical Models Considered

The focus of our simulation study was to compare performance of multiple statistical models for estimating policy impact using annual state-level outcomes, given a policy landscape in which states implemented a given policy at different times. We compare the classic two-way fixed effects DID model to three additional models, selected based both on the previous gun policy simulation study (19) as well as a review of methods commonly used in opioid policy evaluations (1). Specifically, we consider: (1) a "detrended" extension of the classic DID model that includes state-specific linear slopes; (2) a one-period lagged autoregressive (AR) model; and (3) generalized estimating equations (GEE) with an autoregressive correlation structure.

To formalize the setting and inferential goal, we use potential outcomes notation for repeated measures data such that $Y_{it1}$ denotes the potential outcome (e.g., opioid-related mortality rate) for state $i$ ($i = 1, \ldots, 50$) if the policy was in effect at time $t$ while $Y_{it0}$ denotes the potential outcome for state $i$ if the policy was not in effect at time $t$. Thus, each state has two potential outcomes at each time point, representing the outcomes that would be achieved with and without the policy in effect. Our primary treatment effect of interest is $\mathrm{E}[Y_1 - Y_0]$, averaging across both states and times, with each state and each time point equally weighted. Let $A_{it} = \{0,1\}$ denote an indicator for whether or not state $i$ had the policy in effect at time $t$ (where $t = 1, \ldots, T$). Then, $Y_{it}^{obs} = Y_{it1} * A_{it} + Y_{it0} * (1 - A_{it})$ denotes the observed outcome for state $i$ at time $t$ as measured longitudinally for state $i$ over time $t = 1, \ldots, T$.

Essentially, classic DID estimation compares the pre-policy to post-policy change in the treated group to the corresponding pre-period to post-period change in the comparison group. This difference-in-differences provides an estimate of the average policy effect, while controlling for time-invariant differences between treated and untreated states and for time-varying exogenous factors (i.e., those that affect both treated and untreated states equally). The classic DID specification is generally implemented as a two-way fixed effects model that includes both state- and time-fixed effects, expressed as:

$$g\left(Y_{it}^{obs}\right) = \alpha \cdot A_{it} + \boldsymbol{\beta} \cdot \boldsymbol{X}_{it} + \rho_i + \sigma_t + \varepsilon_{it} \tag{1}$$

where $g(.)$ denotes the generalized linear model (GLM) link function (e.g., linear, log), $\boldsymbol{X}_{it}$ denotes a vector of time-varying state-level covariates and $\varepsilon_{it}$ denotes the error term. State fixed effects, $\rho_i$, quantify potential differences in the outcome across states, and time fixed effects, $\sigma_t$, quantify temporal national trends. The coefficient estimate $\hat{\alpha}$ represents the DID estimator, namely the policy effect of $A$ after accounting for differences between states implementing and not implementing a policy and time trends.

Standard DID models assume that the difference in the outcomes of the treated and untreated groups would remain constant in the absence of the policy intervention (with magnitude equal to that observed pre-policy). In practice, this assumption is often referred to as the "parallel trends" assumption, although we note that "parallelism" is actually a stronger assumption than necessary, as trajectories need only be equivalent, not necessarily parallel in the linear sense [23]. The outcome levels themselves are not assumed to be equivalent across groups; level differences are accounted for by the state fixed effects. A common misperception is that this assumption can be tested by assessing whether pre-policy period trends are parallel; however, this assumption is inherently untestable as it involves the unobservable counterfactual trends in the post-period. Indeed, conducting "tests of parallel trends" in the pre-period can lead to bias and misleading results [23].

The second model that we evaluate is an extension of the classic DID model that additionally includes state-specific slopes (referred to as "detrending" the data). The detrended model can be expressed as:

$$g\left(Y_{it}^{obs}\right) = \alpha \cdot A_{it} + \boldsymbol{\beta} \cdot \boldsymbol{X}_{it} + \rho_i + \sigma_t + \sum_{s=1}^{50}(\omega_s \cdot t) + \upsilon_{it} \tag{2}$$

where $\omega_s$ denotes the state-specific linear slope over time and $\upsilon_{it}$ denotes the error term. This model expands on Equation (1) by adding in state-specific linear trends $\left(\omega_s \cdot t \cdot 1(state_i = state_s)\right)$. In this model, each state has its own fixed effect to account for its mean as well as a unique linear slope over time. Because the model also includes a national time trend (fit via year fixed effects), the state-specific linear trend is interpreted as the difference between the national time trend and the state trend. This model may be used as a robustness check to rule out differential state trajectories over time – i.e., if Equations (1) and (2) yield similar policy effects, this suggests the absence of differential trajectories (see Bilinski and Hatfield [23] for a discussion of this approach). In the presence of differential trajectories that are additive, Equation (2) should offer an improvement over Equation (1). However, caution must be used, as the time trend terms may functionally "over control" and absorb part of the treatment effect in addition to pre-existing differential trends, particularly in the presence of a time-varying treatment effect [24].

Additionally, we considered an AR model, as the prior gun policy simulation study found that AR models performed especially well when estimating the policy effect on total firearms deaths [17]. AR models include one or more lagged measures of the outcome (e.g., $Y_{it-1}^{obs}$) as covariates to control for potential average differences in outcome trends across treated and comparison states. These models can improve prediction when outcomes are highly autocorrelated, as is the case with annual measures of state-level opioid-related mortality. The AR model examined here included a single lagged value of the outcome (as this was identified as the top performing AR model in the prior gun policy simulation study), expressed as:

$$g(Y_{it}^{obs}) = \alpha \cdot (A_{it} - A_{i,t-1}) + \boldsymbol{\beta} \cdot \boldsymbol{X}_{it} + \gamma \cdot Y_{it-1}^{obs} + \sigma_t + \epsilon_{it} \tag{3}$$

Akin to Equation (1), this model includes time fixed effects, $\sigma_t$, to quantify temporal trends across time, but adjusts for state-specific variability through the use of the AR term ($\gamma \cdot Y_{it-1}^{obs}$) rather than state fixed effects. Notably, inclusion of the AR term creates a "change" model, as the policy effect is defined as the expected difference in the outcome, given the prior year's outcome. As such, we coded the policy variable ($A$) using *change coding* ($A_{it} - A_{i,t-1}$), based on early work demonstrating that effect size estimates from AR models can be substantially biased when using standard *effect coding* ($A_{it}$) (25). An AR model with a single lagged outcome is very closely related to the first-difference estimator, a commonly-used alternative to the fixed effects estimator (e.g., Equation (1)). Indeed, when there are only 2 time periods, a first-difference estimator and fixed effects estimator are identical; with 3 or more time periods, the relative performance of these estimators depends on the degree of autocorrelation in the outcome (18, 26).

Finally, we considered a fixed effect model using GEE. In the context of correlated outcomes (e.g., within states), GEE model parameters are estimated by specifying a covariance structure for the clustered outcomes (27). This model can be expressed as:

$$g(Y_{it}^{obs}) = \alpha \cdot A_{it} + \boldsymbol{\beta} \cdot \boldsymbol{X}_{it} + \sigma_t + \zeta_{it,} \tag{4}$$

which includes time fixed effects $\sigma_t$ and time-varying state-level confounders measured in $\boldsymbol{X}_{it}$. GEE is a semi-parametric method that requires specification of the covariance matrix for within-subject observations (e.g., exchangeable, autoregressive, unstructured). We assume an autocorrelation structure of order 1 (AR1) which means the correlation structure $\mathbf{R}$ for the repeated measures within each state is

$$R_{t,m} = \begin{cases} 1 \; if \; t = m \\ |\rho^{t-m}| \; if \; t \neq m \end{cases}$$

for the *t, m* element of $\mathbf{R}$.

Overall, in the context of a longitudinal policy evaluation study, the central challenge is disentangling what degree, if any, of the observed heterogeneity in outcomes across states is due to a true policy effect versus other factors. All models we considered included time fixed effects to account for state-invariant (i.e., national) temporal trends. Additionally, the classic DID and detrended DID both included state fixed effects in order to reduce bias due to time-invariant factors that vary across states. In contrast to fixed effects, the autoregressive model adjusts for state-specific variability through the use of the lagged outcome term and a GEE approach uses an AR correlation structure to account for correlation at the state-level. The optimal model should be the one for which the underlying assumptions of the model match the true processes generating the data. As it is impossible to test model assumptions in practice, we used a simulation study with a known data-generating process to assess the relative performance of these statistical models.

### 2.3 Statistical Models Tested via Simulation

Within our four primary DID variations (i.e., classic two-way fixed effect model, detrended model, autoregressive model, and GEE model), we additionally considered three other estimation aspects: GLM link function specification, standard error estimation, and weighting to account for state population. We detail each below and summarize all candidate models in **Table 1**.

**Table 1.** Overview of statistical models evaluated in simulation study

| Regression specification | Link function | SE estimation | Population weighting |
|---|---|---|---|
| Classic 2-way Fixed Effects | Linear | none; Huber; cluster | Population weighted; unweighted |
| | Log-linear | none; Huber; cluster | Population weighted; unweighted |
| | Negative Binomial | none; Huber; cluster | Unweighted, with log(population) used as an offset |
| | Poisson | none; Huber; cluster | Unweighted, with log(population) used as an offset |
| Detrended | Linear | none; Huber; cluster | Population weighted; unweighted |
| | Negative Binomial | none; Huber; cluster | Unweighted, with log(population) used as an offset |
| Autoregressive | Linear | none; Huber; cluster | Population weighted; unweighted |
| | Log-linear | none; Huber; cluster | Population weighted; unweighted |
| | Negative Binomial | none; Huber; cluster | Unweighted, with log(population) used as an offset |
| | Poisson | none; Huber; cluster | Unweighted, with log(population) used as an offset |
| GEE | Linear | AR1 structure | Population weighted; unweighted |

(1) _GLM specifications_: As opioid-related deaths are discrete and historically rare events, count models or models accounting for the skewed nature of the outcome may be more appropriate than traditional linear models assuming normality. We tested the relative performance of the following GLMs: linear, log-linear (a linear model with log-transformed outcome), and two log-link models (negative binomial and Poisson).

(2) _Standard error (SE) estimation:_ There are 3 commonly used ways to estimate the SE of the effect estimate: (1) no adjustment; (2) Huber adjustment: robust estimators (also known as sandwich estimators, or Huber corrected estimates) that attempt to adjust the SE for violations of distributional assumptions (28, 29); and (3) cluster adjustment: adjustments to account for possible violations of the assumed independence of observations within states (28-30). For each model (except the GEE model), we estimated the SE in these three ways. For the GEE models, we used the AR(1) covariance structure for our SE estimation. We also note that we additionally considered the Arellano method (31) as implemented in R's vcovHC package, yet do not report these results for parsimony, as they were very similar to the Huber method (see our Shiny tool for full details).

(3) _Use of state population weights:_ Finally, we explored the impact of using state population as an analytic weight in the linear and log-linear models, an approach commonly used in state-level policy evaluations (e.g., within opioid-related policy studies (32-35)), For state-level analyses of opioid-mortality rates, the use of population weights puts equal weight on each death, regardless of which state it occurred in, whereas unweighted analyses put equal weight on each state, such that a death in a small state will have much greater weight than a death in a larger state. We note that data was generated such that policy effects are constant across all states regardless of size or other characteristics, so weighting is not expected to affect bias but may have substantial effects on the SE estimates. Given that log-link models (e.g., negative binomial, Poisson) are estimated using mortality counts (rather than rates) and do not need to be weighted to be nationally-representative, we did not examine the impact of weighting in these models. Instead, these models include the logarithm of state population size as an offset, resulting in a model that is effectively predicting the opioid-related death rate, such that exponentiated model coefficients can be interpreted as incident risk ratios.

# 3. SIMULATION DETAILS

This section describes our simulation study in detail, including the data sources used in the study, the data generation scheme, and the performance metrics used to compare the approaches.

## 3.1 Data Sources and Measures

The outcome of interest is the annual state-specific opioid mortality rate per 100,000 state residents, obtained from the 1999-2016 National Vital Statistics System (NVSS) Multiple Cause of Death mortality files. Consistent with other studies (36-38), opioid related overdose deaths were identified based on ICD10-CM-external cause of injury codes X40-X44, X60-64, X85, and Y10-Y14, indicating accidental and intentional poisoning, with opioid overdose based on the presence of one of the following diagnosis codes: T40.1 poisoning by heroin, T40.2 poisoning by natural and semisynthetic opioids (e.g., oxycodone, hydrocodone), T40.3 poisoning by methadone, and T40.4 poisoning by synthetic opioids excluding methadone (e.g., fentanyl, tramadol).

Given concerns about model overfitting in the presence of numerous covariates (39), we included only a single covariate: state-level unemployment rate (40). This covariate was selected because of the frequency of its use in opioid policy studies (1). Sensitivity analyses including a broader set of covariates (e.g., poverty rates, income levels, and percentages in defined race/ethnicity and age groups) resulted in no meaningful change to the general findings with a slight increase in precision; as such, we present findings from the more parsimonious model.

## 3.2 Simulation Data Generation

The simulation design builds directly from prior work that compared statistical methods for evaluating the impact of state laws on firearms deaths (19). For each simulation iteration, 5,000 simulated datasets were generated.

In each simulated dataset, a random subset of $k$ states were selected to be the policy/treated group, with remaining states serving as the comparison/untreated. This simulation represents the simplified scenario in which there is no confounding by observed or unobserved covariates or by lagged values of the outcome, $Y_{it-1}^{obs}$. For each state and year, a time-varying indicator $A_{it}$ was generated to denote whether the hypothetical policy was in effect. For comparison states, $A_{it} = 0$ for the entire study period. For policy states, the month and year of policy enactment were randomly generated, with year restricted to 2002-2013 (inclusive) to ensure at least three years of outcome data both before and after enactment. In the first year of implementation, $A_{it}$ was coded as fractional value between 0 and 1, indicating the percentage of the year the policy was in effect. Once a policy was implemented, it remained in effect throughout the study period; thus, $A_{it} = 1$ for all remaining years.

As we were considering models with different log links, we evaluated their performance using simulated data for which each model was correctly specified, so as to facilitate comparison across models. Simulated outcome data were generated as follows: For untreated states, outcome values were set equal to the actual observed state-specific, year-specific opioid overdose rates for all times $t$, namely $Y_{it}^{obs} = Y_{it0}$. Similarly, for treated states in the pre-policy period, values are also equal to the actual observed values ($Y_{it}^{obs} = Y_{it0}$). For treated states in the post-policy period, outcomes $Y_{it1}$ were generated by augmenting the observed value $Y_{it0}$ with an effect size of magnitude $\alpha$ as follows: $Y_{it1} = Y_{it0} + \alpha_{linear}$ for linear models; $Y_{it1} = Y_{it0} + \log(\alpha_{log})$ for log-linear models; and $Y_{it1} = Y_{it0}*(\alpha_{log} - 1)$ for log link models.

Simulation conditions varied the following factors:

(1) _Effect size_. We considered settings when the policy had a null effect, as well as a non-null effect of small, medium and large magnitude. For null effect conditions ($\alpha = 0$), post-policy observations were equal to actual observed values, $Y_{it}^{obs} = Y_{it0}$, for both treatment groups. When generating non-null effects, we tailored the magnitude of $\alpha$ with respect to link function (i.e., $\alpha_{linear}$, $\alpha_{log}$) to ensure that the magnitude of the resulting effect, calculated in terms of the mean number of additional deaths nationally (per 100,000 people), was comparable across models. Specifically, we started by generating data with an $\alpha_{log} = \pm 5\%$ (small), $\pm 15\%$ (medium), and $\pm 25\%$ (large) on the multiplicative scale and then empirically calculated the average excess mortality count across simulated datasets for each effect size. We then specified the corresponding $\alpha$ values for the linear models such that they would yield an effect size of the same magnitude (i.e., $\alpha_{linear} = \pm 0.23$, $\pm 0.70$, and $\pm 1.16$).

(2) _Number of treated units_. We also investigated the role of the number of policy states, simulating data in which 1, 5, 15 and 30 states implemented the policy. Note that the total sample size of treated and untreated states is always 50.

(3) _Timing of policy effect_. State policies often do not become 100% effective immediately after implementation, making it important to consider variation in the onset of policy effectiveness. We considered two possible conditions: an instantaneous effect and a 3-year linear phase-in effect. In both the data generating and analytic models, an instantaneous effect was specified as a simple step-function that has a value of zero when the policy is not in effect and a value of one when the policy is in effect (as described above). The gradual policy effect allows for the effect of the policy to grow linearly in the first 3 years after implementation with values starting at zero and reaching 1 after 3 years of implementation.

### 3.3 Metrics for Assessing Relative Performance of Candidate Statistical Methods

Performance metrics include directional bias, magnitude bias, and root mean squared error, as well as Type I error and rate of correctly rejecting the null hypothesis, given the prevalence of frequentist NHST in the applied literature.

(1) _Directional bias_. Directional bias assesses the average difference between the estimated effect and true effect over all simulations for a given effect size (e.g., $\pm 5\%$), showing the tendency of the estimated effects from a given model to fall closer or further from the true effect on average. We report directional bias summarized over both the positive effect size conditions (e.g., $+5\%$) as well as the negative effect size conditions (e.g., $-5\%$) to quantify how the models are doing on average for a fixed effect size $\alpha$, regardless of the direction. We define directional bias as the average of the sum of the bias across positive and negative effect simulations, as follows:

$$DirectionalBias_\alpha = \left( \sum_{k=1}^{5000} \frac{\widehat{\alpha}_{k,pos} - \alpha_{pos}}{5000} + \sum_{k=1}^{5000} \frac{\widehat{\alpha}_{k,neg} - \alpha_{neg}}{5000} \right) / 2$$

Additionally, we standardized bias by reporting it with respect to the mortality count for both linear and nonlinear models to facilitate comparison across models. Then, we converted the standardized directional bias into percent directional bias by dividing it by the expected change in mortality count that corresponds to the given $\alpha$ (e.g., when $\alpha = \pm 5\%$ the expected change in deaths nationally will equal $\pm 700$, respectively).

(2) _Magnitude bias._ Magnitude bias assesses whether the estimated effects are systematically too small or too large, relative to the true effect. Magnitude bias is computed by taking the average of the bias across the positive and negative effect simulations, after multiplying the bias from the negative effect simulations by negative one.

$$MagnitudeBias_\alpha = \left(\sum_{k=1}^{5000}\frac{\hat{\alpha}_{k,pos}-\alpha_{pos}}{5000} - \sum_{k=1}^{5000}\frac{\hat{\alpha}_{k,neg}-\alpha_{neg}}{5000}\right)/2$$

For example, with a model that shows a magnitude bias of +0.1 with a true effect size of ±0.30, the model typically gives estimates of +0.4 or –0.4 for the positive and negative effect versions of the simulation, respectively, exaggerating the true effect size in both cases. Conversely, a model that shows a magnitude bias of -0.1 would give estimates of +0.3 or –0.2 for the positive and negative effect simulation, respectively, underestimating the true effect size. As with directional bias, we standardized magnitude bias so it represents mortality count and report percent magnitude bias below by dividing it by the corresponding expected change in deaths nationally that would correspond to the given $\alpha$.

(3) _Root mean squared error (RMSE)_. RMSE is calculated by taking the square root of the sum of the mean squared errors (e.g., $\sqrt{\sum_{k=1}^{5000}(\hat{\alpha}_k - \alpha)^2/5000}$ ). RMSE quantifies error for a given model specification, taking into account both directional bias and variance.

(4) _Type I error rate_. In the context of traditional NHST, Type I error rate is the frequency of incorrectly rejecting the null hypothesis (i.e., there truly is no policy effect). When data are generated such that there is no true policy effect (i.e., the null hypothesis is true), the model should identify a statistically significant effect (i.e., reject the null hypothesis) no more than 5% of the time if tested with an 0.05 level of significance.

(5) _Correct NHST rejection rates_. We also assessed the ability of the model to correctly identify that the null hypothesis is false in the context of traditional NHST. We quantify the "rate of correct rejections" for each model by calculating the proportion of estimates that were both statistically significant and in the same direction as the true effect. When conducting this significance test, we used a SE correction factor to ensure comparability of correct NHST rejection rates across models with the exact same Type I error rate. Without applying the SE correction factor, models that underestimate the true error in their estimates would appear to have excellent statistical correct rejection rates, even though the actual sampling variability in their estimates may be quite high, in which case the model may not actually be sensitive to detecting a true effect. Typically, analyses are considered to have adequate statistical correct rejection rates/power if the likelihood that they correctly reject the null hypothesis is 80% or higher.

Simulations were conducted in R using the forthcoming OPTIC.simRM library; code is available in the appendix. Extensive results for all statistical models considered in our simulation are available via a Shiny tool (https://elizabethmcneer.shinyapps.io/statmodelsim/).

## 4. RESULTS

In each section below, we first compare results for the set of four linear models (i.e., linear two-way fixed effects, linear detrended, linear AR, and linear GEE models). We then discuss the relative performance across different GLMs (i.e., negative binomial, Poisson, and log-linear models). For parsimony, all summary statistics are averaged across simulation conditions with a gradual policy effect and an instantaneous policy effect.

### 4.1. Directional bias

**Figure 1** shows percent directional bias as a function of both effect size magnitude and the number of policy states for the four different linear models (using population weights). In all cases, percent directional bias decreased both as effect size increased and the number of policy states increased. Most notably, the linear two-way fixed effects model (the classic DID model) had high percent directional bias when the number of treated states was lower than 15 (e.g., ranged from 22% to 291%) (**Figure 1a**). The linear GEE had similar directional bias to the linear two-way fixed effects (ranged from 0% to 305%) (**Figure 1d**). Directional bias was much lower for the detrended model and AR models compared to the two-way fixed effects and GEE models (ranging from ±3% to -21%) (**Figures 1b and c**).

 **Figure 2** shows the percent directional bias for all models under the small effect size condition. Notably, the majority of models had positive directional bias suggesting estimated effects tend to be numerically larger in a positive way on average, regardless of the direction of the true policy effect. The large majority of models had very high rates of directional bias. For example, non-linear models yielded directional bias ranging from 64% to 162%, which translates into excess mortality estimates that are off by 448 to 1,134 more deaths. Directional bias was smallest in the linear models (ranging from -2% to -12%), with the exception of the weighted linear two-way fixed effects and weighted GEE models where directional bias was quite large (116% and 109%, respectively).

 The directional bias was relatively similar between weighted and unweighted versions of both the linear AR and linear detrended models. In contrast, directional bias was significantly larger for weighted version, compared to the unweighted version, for both the traditional DID model (unweighted=-2%; weighted=109%) and linear GEE model (unweighted=-3%; weighted=116%). Further, directional bias was notably larger when there was a gradual versus an instantaneous policy effect, although the magnitude of this difference varied by model.

### 4.2. Magnitude bias

 Broadly, as seen with directional bias, magnitude bias decreased as both effect size and number of policy states increased. We present magnitude bias results for all models under the small effect size condition (**Figure 3**). Magnitude bias was less than 10% for most models, with the exception of the four non-linear AR models (14-25% for the negative binomial, Poisson, and log-linear AR models). Most of the models with non-zero magnitude bias had positive magnitude bias (i.e., overestimating the true policy effect), ranging from 4% (negative binomial 2-way fixed effects and detrended models) to 25% (Poisson AR model). In contrast, the linear AR model had negative magnitude bias (i.e., underestimating the true policy effect), ranging from -4% (with population weights) to -2% (no population weights). For each GLM type, magnitude bias was greater for the AR model compared to the two-way fixed effect or detrended models.

 The use of population weights in the linear and log-linear models did not consistently or notably influence magnitude bias. Furthermore, the magnitude bias remained essentially 0% for the linear two-way fixed effects, linear GEE, and linear detrended models for both the gradual and instantaneous policy effect conditions. For all the other models, magnitude bias was consistently higher for the gradual versus the instantaneous effect conditions (e.g., for the negative binomial AR model, magnitude bias was 10% for the instantaneous condition and 23% for the gradual condition).

### 4.3 Root mean square error

 **Figure 4** shows the average RMSE for simulation conditions with a null treatment effect. Among linear models, AR models had the lowest RMSE (1.08-1.12) compared to the two-way fixed effects models (1.67-1.78), detrended models (1.63-1.69), and GEE models (1.37-1.92) (**Figure 4a**). For the two-way fixed effects, detrended, and GEE models, RMSE was lower for the unweighted models than the corresponding weighted models; however, for the AR models, population weighting yielded slightly lower RMSE. Among non-linear models, the negative binomial models had consistently lower RMSE compared to the Poisson and log-linear models (**Figure 4b**). For the negative binomial model, the detrended and two-way fixed effects models had the lowest RMSE (0.22) while the AR model had the highest RMSE (0.31). Finally, as expected, RMSE was larger for simulation conditions with a gradual policy effect relative to an instantaneous effect (e.g., for the linear population two-way fixed effects model, RMSE=1.58 for instantaneous and RMSE=1.95 for gradual).

### 4.4 Type I error rates

**Figure 5** presents the Type I error rates for the four linear models (using population weights). Type I error rates were very high for the classic DID two-way fixed effects model (**Figure 5a**), ranging up to 67%. Cluster SE adjustment greatly reduced the Type I error rates for this model when 5 or more states implemented a policy, but they were still 2 to 3 times larger than the traditional target of 5%, ranging from 9% to 17%. The detrended model (**Figure 5b**) generally had slightly lower Type I error rates than the two-way fixed effects model, with Type I error rates mostly less than 40%. Notably, the AR model (**Figure 5c**) did not require use of any SE adjustment to obtain appropriate Type I error rates for conditions with 5 or greater policy states (e.g., Type I error rates ranged from 4% to 6%); in fact, SE adjustments in the AR models tended to inflate the Type I error rates. For linear GEE models (**Figure 5d**), Type I error rates were 18% or less for simulation conditions with at least 5 policy states, though rates were still 2-3 times higher than the traditional target of 5%. As in the case of linear models, AR models performed best, followed by detrended models, then two-way fixed effects models in the case of non-linear models (log-linear, Poisson, and negative binomial).

For linear models, population weighting yielded slightly higher Type I error rates for the two-way fixed effects, detrended, and GEE models compared to the corresponding unweighted models (see Shiny Application). In contrast, for the AR models, population weighted models did not consistently perform better or worse than unweighted models. Additionally, Type I error rates were higher (by approximately 8 percentage points) for simulation conditions with a gradual relative to an instantaneous effect.

Given the top performance of the AR model, we also present the relative performance of the AR model across four different GLMs: linear (unweighted), log-linear (unweighted), Poisson, and negative binomial (**Figure 6**). Similar to the results seen for the linear AR weighted model (Figure 4), very good Type I error rates are obtained in the absence of SE adjustment for linear AR unweighted model, the log-linear AR unweighted model, and the negative binomial AR model, regardless of the number of policy states. We note that this does not hold for the Poisson AR model.

### 4.5 Correct NHST rejection rates

**Figure 7** shows correct NHST rejection rates as a function of both the effect size and the number of policy states for the linear models (using population weights). In all cases, as expected, correct rejection rates increased both as the effect size increased and the number of policy states increased, with maximum values obtained for the simulation condition with 30 policy states and a large effect size. For the two-way fixed effects model (**Figure 7a**), correct rejection rates were low across all effect sizes, with a maximum value of 27%. In contrast, correct rejection rates were highest for the AR model (**Figure 7c**), which achieved a maximum value of 73% (nearly the desired 80% rate). Relative to the two-way fixed effects model, correct rejection rates were similar for the GEE model (maximum value=30%) and slightly higher for the detrended model (maximum value=41%). Importantly, all models considered had extremely low correct rejection rates for simulation conditions with a small effect size – e.g., the rate of correctly rejecting the null hypothesis was 8% for negative binomial models and ranged from 4% to 11% across linear models.

For linear and log-linear models, correct rejection rates tended to be higher for unweighted models relative to weighted models. Specifically, the linear two-way fixed effects model yielded a correct rejection rate of 40% for the unweighted model compared to 27% for the unweighted model for the simulation condition with 30 policy states and a large effect size. Similarly, the unweighted linear AR model yielded the correct rejection rate of 81% (compared to 72% for weighted) and the unweighted GEE model yielded the correct rejection rate of 67% (compared to 30% for weighted). Correct rejection rates were consistently smaller (by 3 percentage points on average) for simulation conditions with a gradual relative to an instantaneous policy effect.

**Figure 8** presents correct rejection rates averaged across all simulation conditions in order to highlight relative performance across models. Correct rejection rates were low across all models but were highest for linear AR models (ranging from 22% to 24%) and negative binomial models (ranging from 20% to 23%). The worst performing models were the linear and log-linear two-way fixed effects models and the linear weighted GEE model (correct rejection rates ranged from 9% to 11%). Correct rejection rates for the Poisson models ranged from 12% (two-way fixed effects model) to 18% (AR model); we note that for all specification, the Poisson model was outperformed by the corresponding negative binomial model.

## 5. DISCUSSION

State-level policy evaluations commonly employ a DID study design; yet model specification varies notably across studies and the field lacks clear guidance on which models are optimal. We conducted a novel simulation study to compare the relative performance of multiple variations of the two-way fixed effect model traditionally used for DID, using simulated data based on actual national opioid mortality data so as to mirror data features encountered in practice. Specifically, we compared the classic, linear two-way fixed effects DID model to three alternative models: a detrended model, an AR model, and fixed effect model estimated with GEE with an AR correlation structure. Within these classes of models, we additionally compared link function specifications, SE estimation methods, and the use of population weighting. As discussed further below, we found that the linear AR model was optimal when the outcome was specified as a mortality rate and a negative binomial model was optimal when the outcome was specified as a mortality count. Despite being widely used in applied research, our results highlighted that two widely-used linear DID models – two-way fixed effect and detrended – were consistently outperformed by the less commonly-used AR linear model, which was consistently optimal in terms of directional bias, RMSE, Type I error, and power. As such, we urge applied researchers to move beyond the classic linear two-way fixed effect DID paradigm and consider the use of AR models. Overall, our results indicated notable differences in the performance of the models considered, which has substantial implications for the conduct and interpretation of state-level policy evaluations.

Results from the present study are highly consistent with findings from a prior gun policy simulation study (19), as both studies identified autoregressive models as a top performing model for estimating state-level policy effects. Given the consistency of these findings, it is likely that advantages of AR models over may generalize contexts beyond opioid- and firearm-related mortality. The present study considers a broader range of simulation conditions than the prior gun policy study (e.g., a range of policy effect sizes (5% to 25%) compared to a single effect size (3%)), which similarly strengthens the generalizability of the results. However the optimal choice of the link function may vary by the characteristics of the outcome variable: the gun policy simulations study, which examined firearm-related mortality, found that the negative binomial AR model was optimal whereas the current study, which examined opioid-related mortality, identified the linear AR models as optimal. Indeed, the negative binomial AR model yielded much higher directional and magnitude bias (relative to the linear AR model), likely due to the greater relative skew in the distribution of state-level opioid-related deaths compared to firearm-related deaths. This suggests there is a benefit to running these types of simulations on specific outcomes to ensure selection of the final optimal model for a given outcome. We have an R library for executing these simulations on any repeated measures levels data (OPTIC.simRM).

We make recommendations for practice in Table 2. Although many of these results have been found by others, they have not been well appreciated in the statistical or applied literature, and questions have remained regarding best practices with real-world data like opioid-related mortality rates. For example, with regard to standard error corrections, prior simulation studies (13, 16) show that cluster adjustments are needed to reduce Type I error rates. Bertrand, Duflo (14) showed that the classic sandwich estimator does poorly with small samples; that paper also shows DID without adjustment has high Type I errors (approximately 45%) in their case study data where they randomly simulated random "placebo" laws, as done here. Our work extends prior work by highlighting the challenges specific to the context of evaluating state-level opioid policies with respect to opioid-related mortality, a widely-used outcome in the field.

**Table 2. Key Takeaways for the Practice**

| |
|---|
| When modeling opioid-related mortality as a crude rate in a linear model inclusion of an autoregressive term significantly improves estimation performance with regard to RMSE. |
| When modeling counts of opioid-related mortality, a negative binomial model performs better than a Poisson model. |
| Linear AR models performed optimally with respect to bias, RMSE, Type I error, and correct rejection rates in the context of estimating state-level policy effects of opioid-related mortality |
| Sample size matters for SE estimation. For linear and log-linear models, clustered SEs significantly improved estimation when the treated group comprised 15+ states, yet they had worse performance than unadjusted SEs in the case of only a single treated state. |

Furthermore, researchers and policymakers must recognize the inherent implications of a fundamentally limited sample size of 50 states (of which perhaps only a few, or even a single state implemented the policy of interest) regarding continued reliance on p-values to determine statistical significance. Under traditional NHST, correct rejection rates for the majority of scenarios was extremely low, lower than 25% across all scenarios considered and only above 50% for the best performing models and when there was a large effect size (25%) and the most balanced allocation to treatment versus control. Additionally, Type I error rates for the majority of models relying on NHST when fewer than 15 states are implementing a new policy were unreasonably high, meaning these models could yield a significant effect estimate when in fact such an effect does not exist. It is critical that researchers use models that minimize Type I error rates whenever possible; use of standard error corrections to ensure a Type I error rate of 0.05 are needed in this context when performing NHST. However, we highly recommend the field overall move beyond traditional NHST, given concerns across a range of scientific areas regarding the use of often arbitrary p-value thresholds within that framework (41). Over-reliance on such tests can lead researchers to miss detecting an effective policy by making a meaningful policy effect not "statistically significant."

Critically, the applied field of state-policy research is still implementing traditional NHST, in spite of the repeated calls from the field of statistics (41) to move beyond reliance on decisions based on whether one has p-values less than 0.05. All of the studies in our recent opioid literature review (1) relied on traditional NHST to determine if their findings on the primary policy were "statistically significant." One alternative approach that holds promise is the use of Bayesian approaches to estimate state-level policy effects. Bayesian methods can be used to estimate effects that directly correspond to the likely effects of the yes/no decisions facing policymakers considering such legislation (namely, the probability that a given law is associated with an increase or a decrease in firearms death), and can also more accurately reflect the large amount of uncertainty in these analyses. For an illustration of an Bayesian approach in context of gun policy (17).

We note that our data generating process only generated synthetic observations for the treated states in the post-period (in order to induce a policy effect of a known magnitude), rather than generating complete trajectories for both treated and untreated states. As such, we (like applied researchers) were not privy to the "truth" about whether the parallel counterfactual trend assumption, the core identifying DID assumption, was upheld; however, since our treated and control states were selected randomly, we do not expect these groups to exhibit systematically differential trajectories. We highlight that the parallel counterfactual trends assumption is untestable, given that this assumption pertains to unobservable counterfactual outcomes. Yet in practice, researchers often conduct a so-called "partial test of parallel trends" by statistically testing whether the pre-intervention trends differ across groups (6, 7). We discourage this practice, as it is not informative regarding the actual underlying counterfactuals and indeed may induce a false sense of confidence in the validity of the common trends assumption. Additionally, a detrended model may be used as a robustness check; if the classic two-way fixed effect model and a detrended model that allows for differential state trajectories over time yield similar policy effects, this provides some evidence in favor of the common trends assumption. See Bilinski and Hatfield (23) and Rambachan and Roth (42) for further discussion of these issues and alternative strategies for assessing plausibility of the parallel counterfactual trends assumption. We also note that if the parallel counterfactual trends assumption holds on one model scale (e.g. linear) it may not automatically hold on other scales (e.g., count). Finally, we highlight that an understanding of state policy environments is also key to assessing whether common trends is a reasonable assumption. In particular, applied researchers should have familiarity with the substantive area, including other policies that states may have enacted during the study period that would be expected to additionally impact the outcomes (see Schuler, Griffin (18) for further discussion).

Fundamentally, longitudinal and panel data do not conform to the traditional regression assumption of independent and identically-distributed (*iid*) residuals. When considering various modeling approaches, it may be helpful to distinguish between three distinct phenomena that contribute to departures from *iid* residuals and to have diagnostic checks for which deviation might be occurring in a given data set: outcome autocorrelation, clustering at the state-level, and departures from model distributional assumptions. First, some degree of autocorrelation in the outcome timeseries is likely. Our results from both the current simulation, as well as the prior gun policy simulation, highlight that cause-specific mortality outcomes are likely to highly autocorrelated. Similarly, autocorrelation is expected for other key health policy outcomes, such as disease-specific incidence rates and healthcare spending measures. The presence of autocorrelation following an AR1 structure can be assessed using the Durbin-Watson test; more generally, an autocorrelation function (ACF) plot, also called a correlogram, can be used to assess the degree of autocorrelation across lagged time periods (Friendly 2002, Durbin and Watson 1971). Autocorrelation is effectively addressed through the use of an AR model or GEE with an AR correlation structure. See Beard, Marsden (43) for a pragmatic discussion of timeseries data analysis in the context of addition research. With regard to state-level clustering, one can compare cluster adjusted versus unadjusted standard errors or compute intracluster correlation coefficients (ICC) to understand how strong the impact of clustering will have on the study design. Though, sample size is a key consideration and such diagnostics like ICCs are not reliable when sample sizes are less than 30 (44). Our results indicate that when in the context of only a single treated state, cluster and Huber SE adjustments yield worse performance than no adjustment. While this has been previously demonstrated in the literature Bertrand, Duflo (14), these insights are often not reflected in the applied literature.

The simulation design has several limitations and future research is needed to build upon this work. First, by randomly selecting states to enact a given policy, this simulation represents the simplified scenario in which there is no confounding by observed or unobserved covariates (including lagged values of the outcome). Future simulation work will consider more complex scenarios, including where such confounding exists given the likelihood that states implementing certain policies differ from states that do not. A growing set of methods aim to deal with potential confounding and need to be considered, including: incorporation of propensity score weighting into the DID framework (45), synthetic control methods (46-48) and augmented synthetic control methods (49), and doubly-robust DID estimators (50), as well as DID extensions that are robust to violations of the parallel trends assumption (51). More broadly, our simulation study did not exhaustively compare models used in practice: for example, we did not consider random effect models in this study, as prior work indicated that they are not commonly used in practice in opioid policy evaluations (1). Second, while the timing of policy enactment varied across treated states, our simulated data had a constant policy effect across states and across time, which may be an unlikely assumption in some contexts. Recent work has showed that in the presence of heterogeneity in policy timing and treatment effects, the classic linear two-way fixed effect DID model yields biased treatment effect estimates (11, 52, 53). Future work is needed to investigate relative model performance in the context of treatment heterogeneity. Finally, while there are numerous outcomes of interest when evaluating the impact of an opioid policy, we focused on fatal overdoses given that approximately 1/3 of published opioid policy evaluation studies examined this outcome. It is unclear how well the results generalize to other opioid or non-opioid outcomes. Future work should entail careful consideration of additional outcomes and extend this line of simulation research to identify optimal model specifications in other policy contexts.

More broadly, as noted by Schell, Griffin (19): "A scientific field built on studies with such low power (e.g., less than 0.20) will have a large fraction of significant results that are spurious, a substantial proportion of significant effects that are in the wrong direction, and significant effects that substantially overestimate the true effect size (54)." There is an urgent need for the field to develop more robust and powerful methods that can be used to help guide state policy. This call is needed to address current public health crises in the U.S. (e.g., opioid epidemic, gun violence, COVID-19) but also extends beyond to future crises that will develop (e.g., climate change). We have to do a better job advancing new approaches that improve accuracy while acknowledge uncertainty in state-level policy effects. Research in these areas is needed to help us ensure we are meeting the needs of applied policy researchers and key decision makers.

## CONCLUSIONS

The findings highlight notable limitations of commonly used statistical models for DID designs, designs widely used in opioid policy studies and in state policy evaluations more broadly. In contrast, the optimal model identified (the AR model) is rarely utilized in state-policy evaluation. We urge applied researchers to move beyond the classic DID paradigm and adopt the use of AR models.

## LIST OF ABBREVIATIONS

| | |
|---|---|
| AR | autoregressive |
| AR(1) | autocorrelation structure of order 1 |
| DID | difference-in differences |
| GEE | generalized estimating equations |
| GLM | generalized linear model |
| NHST | null hypothesis significance testing |
| NVSS | National Vital Statistics System |
| RMSE | root mean squared error |
| SE | standard error |

## DECLARATIONS

**Ethics approval and consent to participate**

The RAND Corporation Institutional Review Board deemed this study exempt (Human Subjects Assurance Number 00003425 (6/22/2023)).

**Consent for publication**

Not applicable.

**Availability of data and materials**

The data that support the findings of this study are available from National Vital Statistics System (NVSS) Multiple Cause of Death mortality files (1989 through 2018) but restrictions apply to the availability of these data, which were used under license for the current study, and so are not publicly available. The data can be request under a similar license (data use agreement) from the NVSS.

**Competing interests**

The authors declare that they have no competing interests.

**Funding**

This research was financially supported through a National Institutes of Health (NIH) grant (P50DA046351) to RAND (PI: Stein). NIH had no role in the design of the study, analysis, and interpretation of data nor in writing the manuscript.

**Authors' contributions**

The authors jointly conceived the idea for the study. BG and TS designed, developed and implemented the original simulation code; BG lead all adaptions and runs under consultation with all authors. EM developed needed graphics and the associated Shiny app for this work. BG, MS, and ES drafted the manuscript with input from all authors. All authors extensively edited and provided input on all phases of the study and all authors read and approved the final manuscript.

**Acknowledgements:** The authors would like to thank Kosali Simon and the participants of the 2019 ASHEcon session that provided helpful comments on an earlier draft. Finally, the authors want to thank Hilary Peterson for her assistance with manuscript preparation and submission.

**FIGURE LEGEND**

**Figure 1.** *Percent directional bias for the four different linear models considered, all with population weights: (1a) the two-way fixed effects model, (1b), the detrended model, (1c) the AR model, and (1d) the GEE model.*

**Figure 2.** *Percent directional bias for all models considered in settings with small effect sizes.* Note: AR = autoregressive, FE = fixed effects, GEE = generalized estimating equation

**Figure 3.** *Percent magnitude bias for all models considered in settings with small effect sizes.* Note: Results showing very small grey line at 0 are equal to 0. The statistics shown will slightly favor linear over non-linear models since we have to convert magnitude bias into a total count of deaths. When magnitude bias measures are converted into the native units of the negative binomial models (log risk ratios), the negative binomial models tended to show slightly better performance relative to the linear models (as seen here).

**Figure 4.** *Root mean squared error for (4a) the linear and (4b) nonlinear models under the null effect simulation condition.* We present this graph stratified by linear and non-linear models, as there is no method to compare RMSE across linear and nonlinear models that yields a fair comparison.

**Figure 5.** *Type I error rates for linear model specifications: (5a) the two-way fixed effects model, (5b), the detrended model, (5c) the AR model, and (5d) the GEE model. Horizontal line denotes the target Type I error rate value of 0.05.*

**Figure 6.** *Type I error rates for the AR models for four different GLMs: (6a) linear (unweighted), (6b) log linear (unweighted), (6c) Poisson, and (6d) negative binomial.* Horizontal line denotes the target Type I error rate value of 0.05.

**Figure 7.** *Correct NHST rejection rates as a function effect size and number of policy states for linear models: (7a) two-way fixed effects DID model, (7b), detrended DID model, (7c) AR model, and (7d) GEE model.* Note: All models were fit with population weights

**Figure 8.** *Average power across all simulation conditions for all models considered in this simulation.*

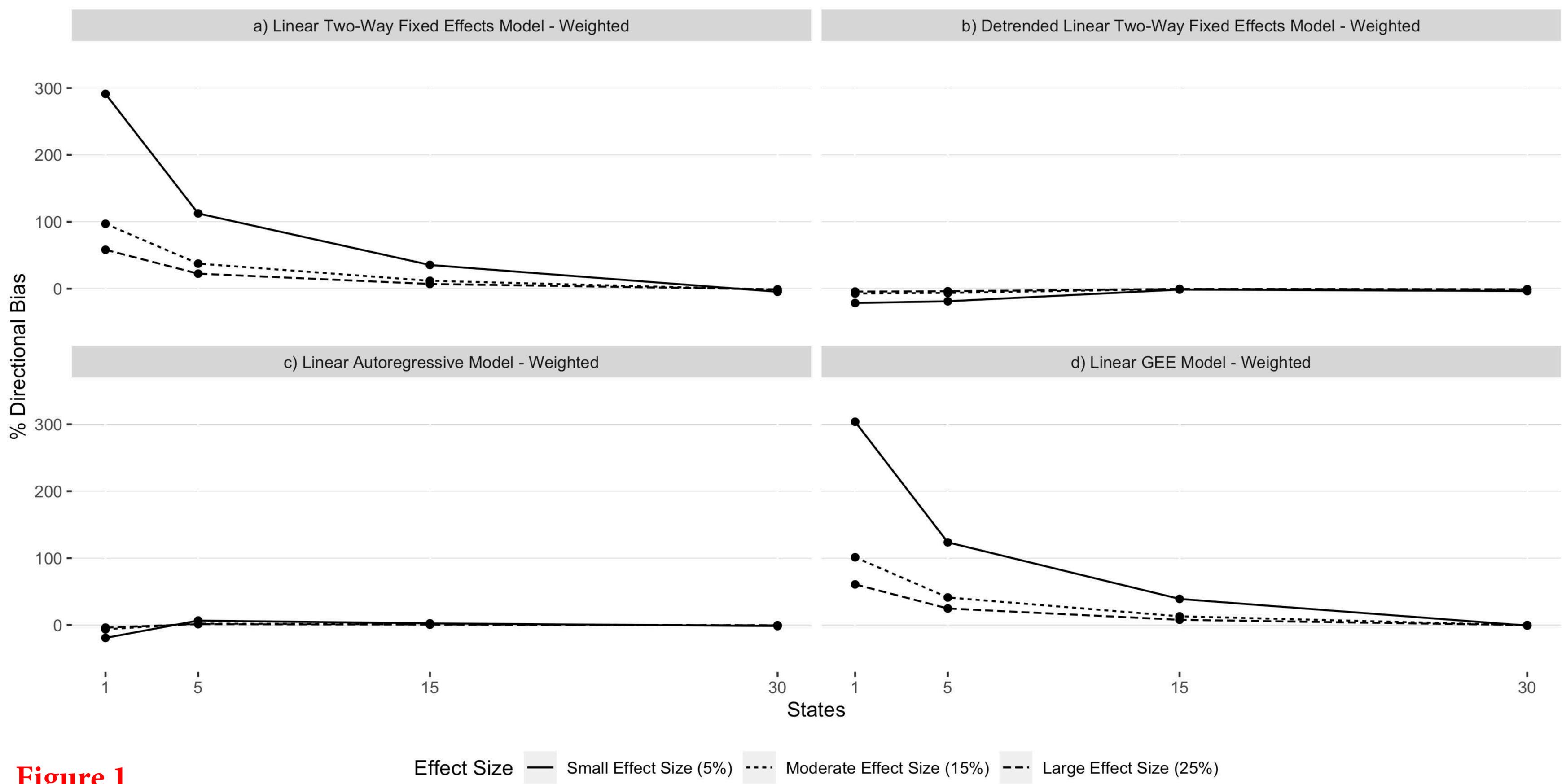

**Figure 1**

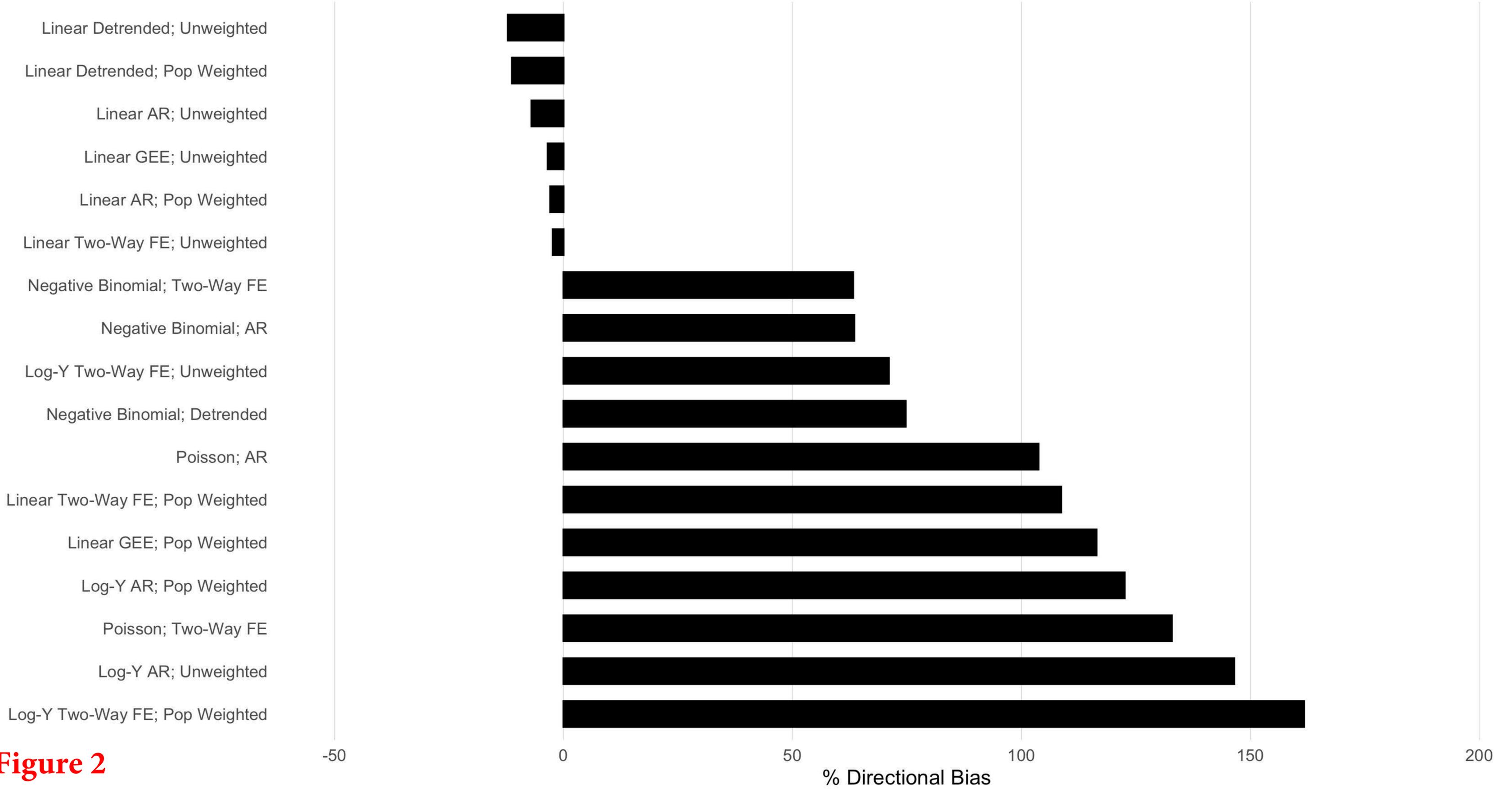

**Figure 2**

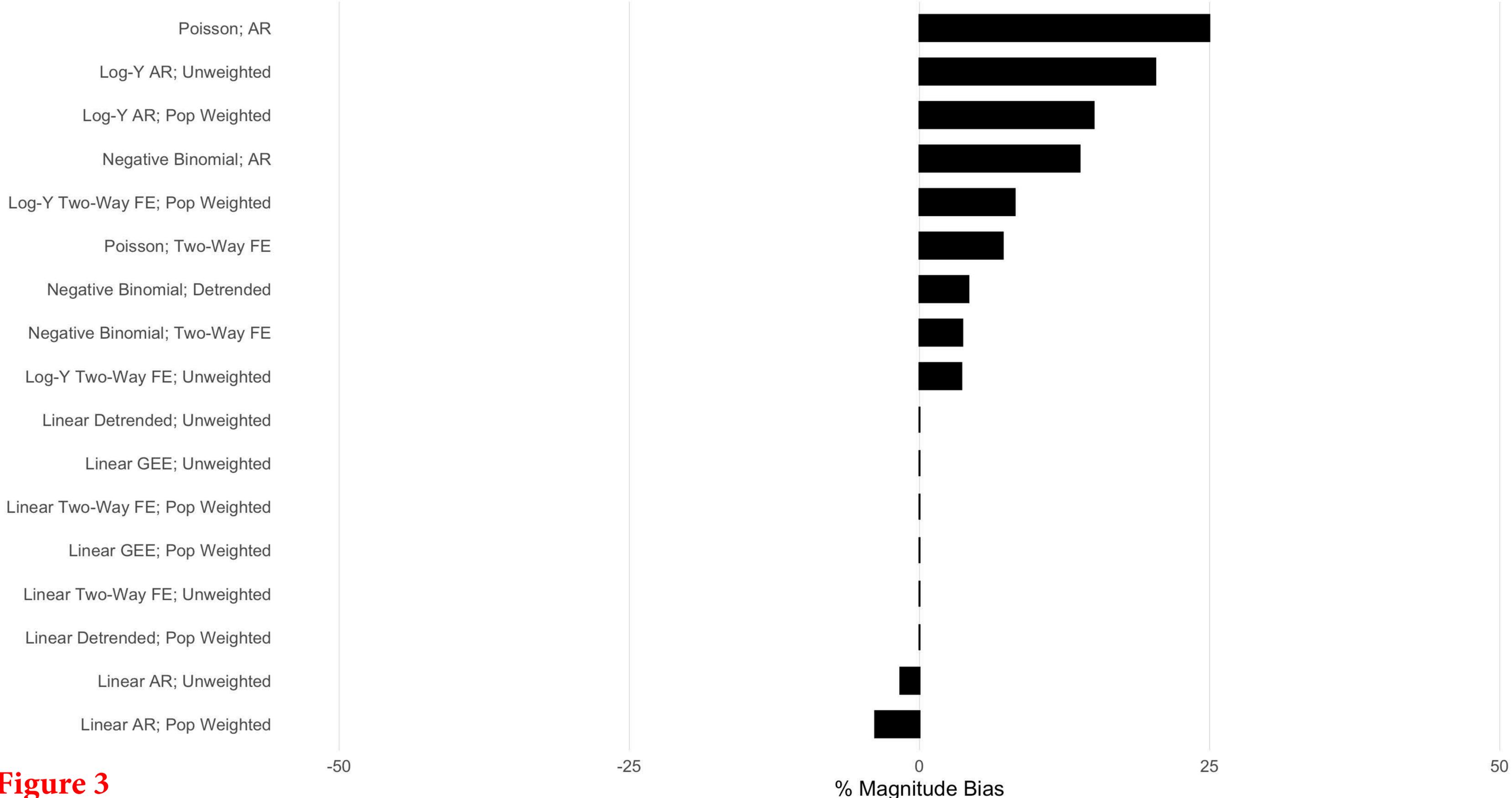

**Figure 3**

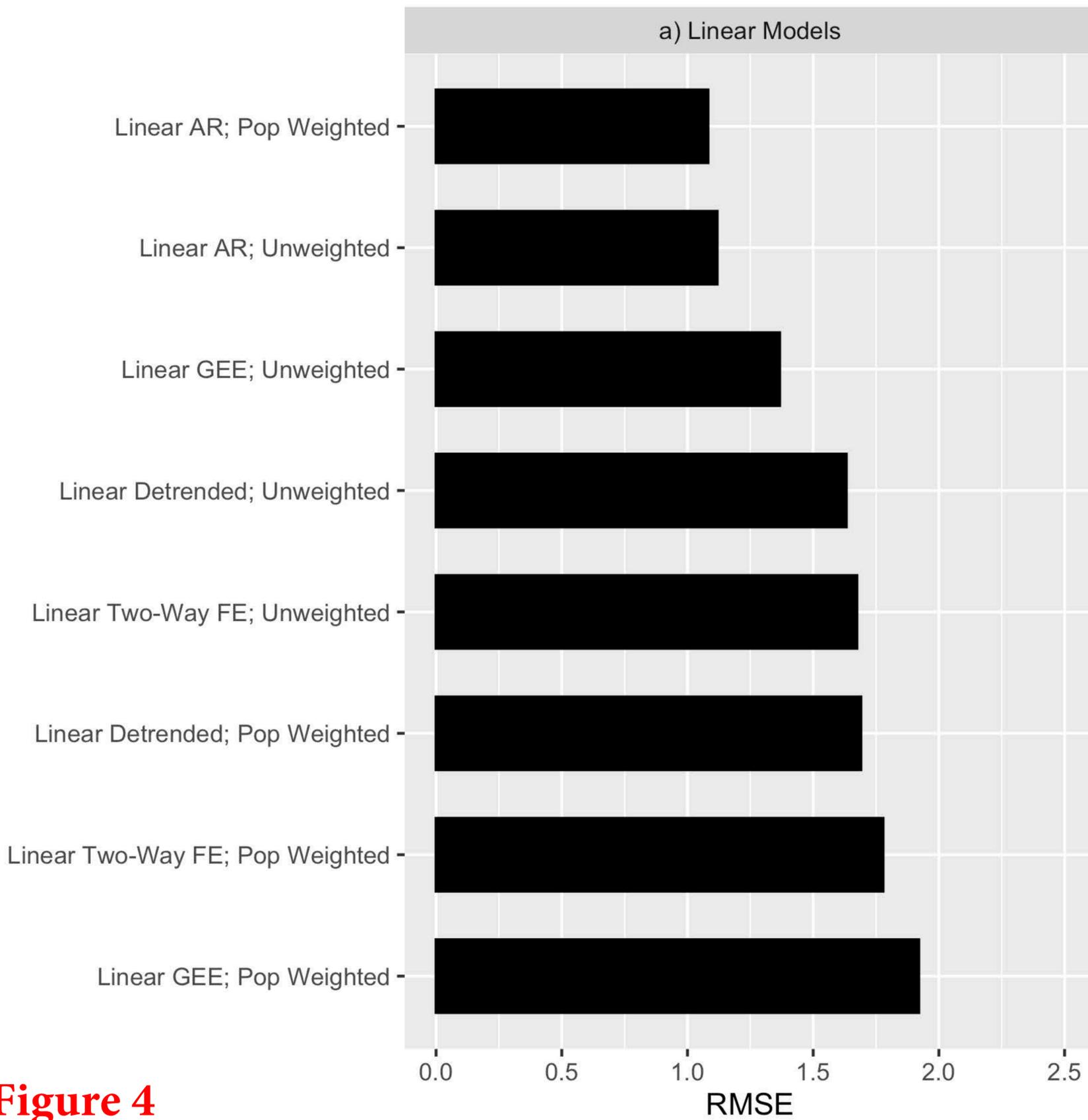
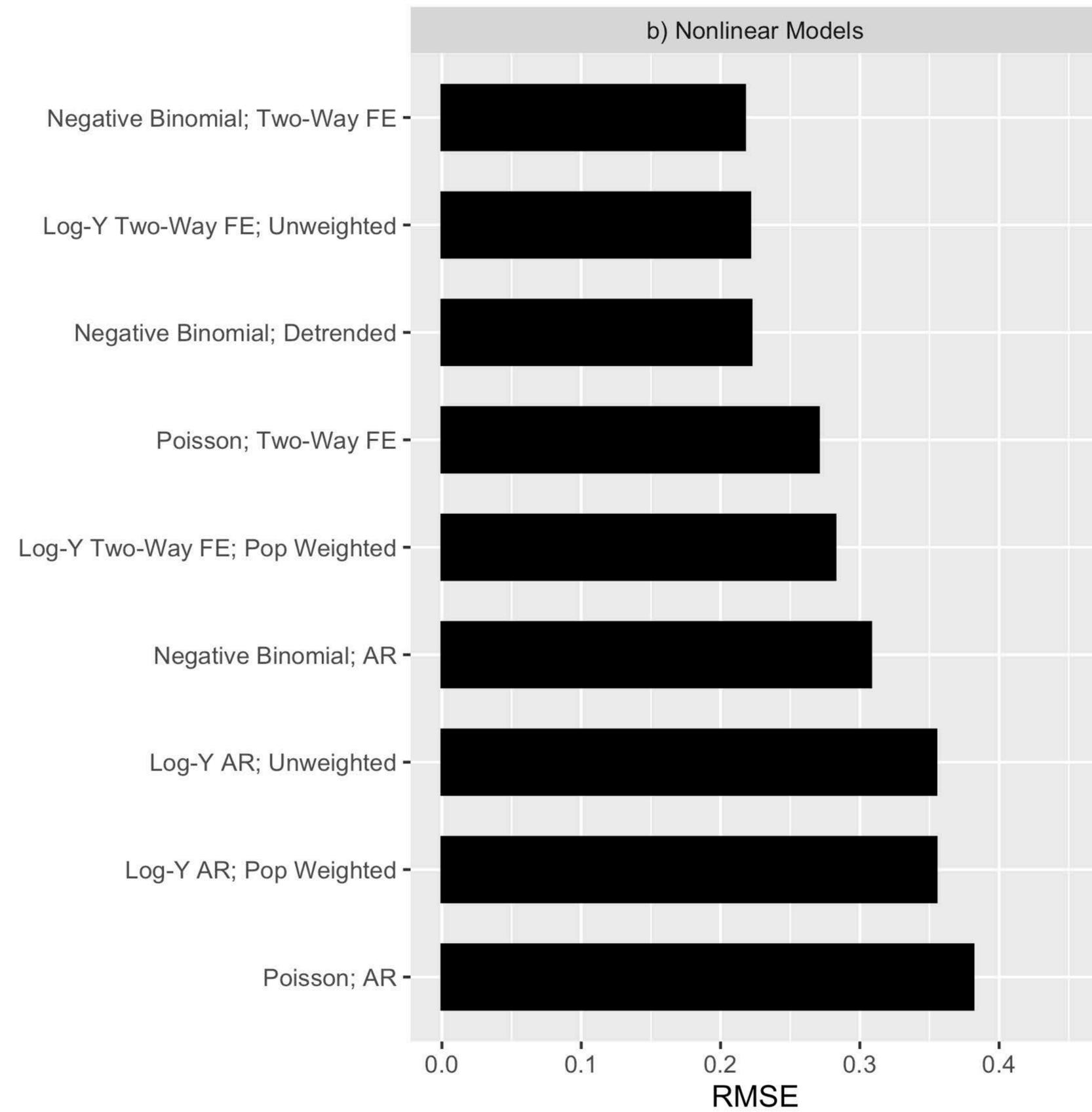

**Figure 4**

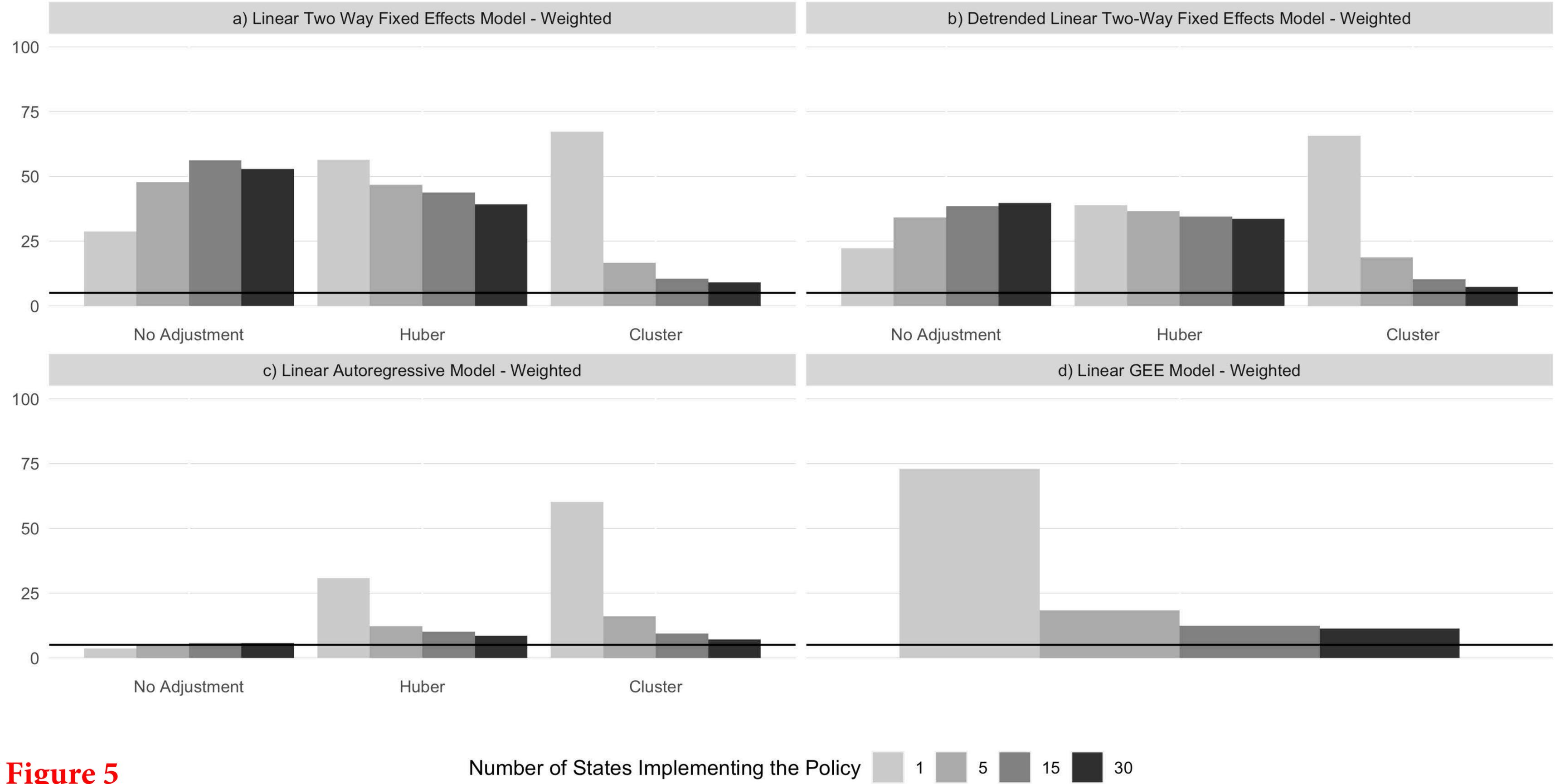

**Figure 5**

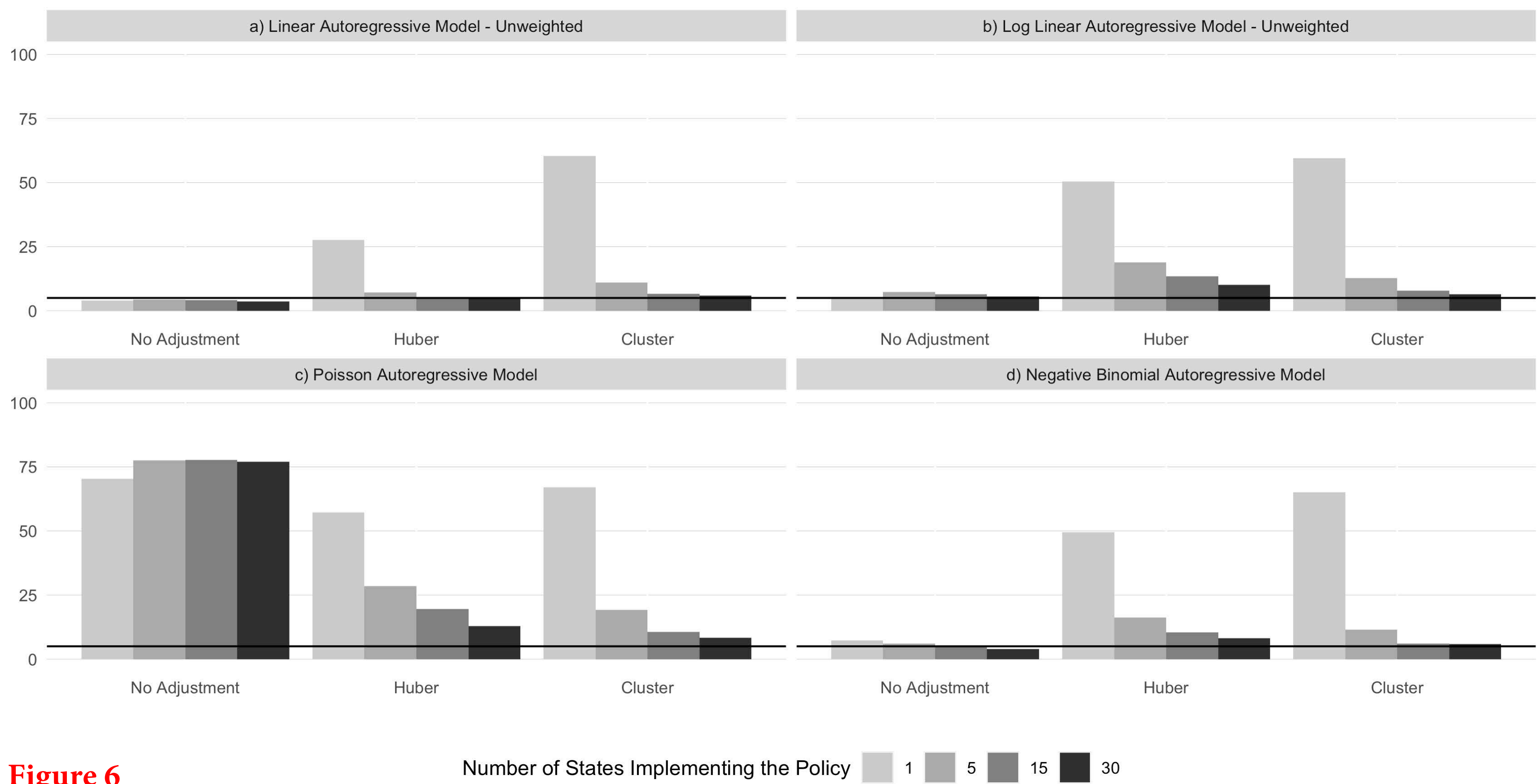

**Figure 6**

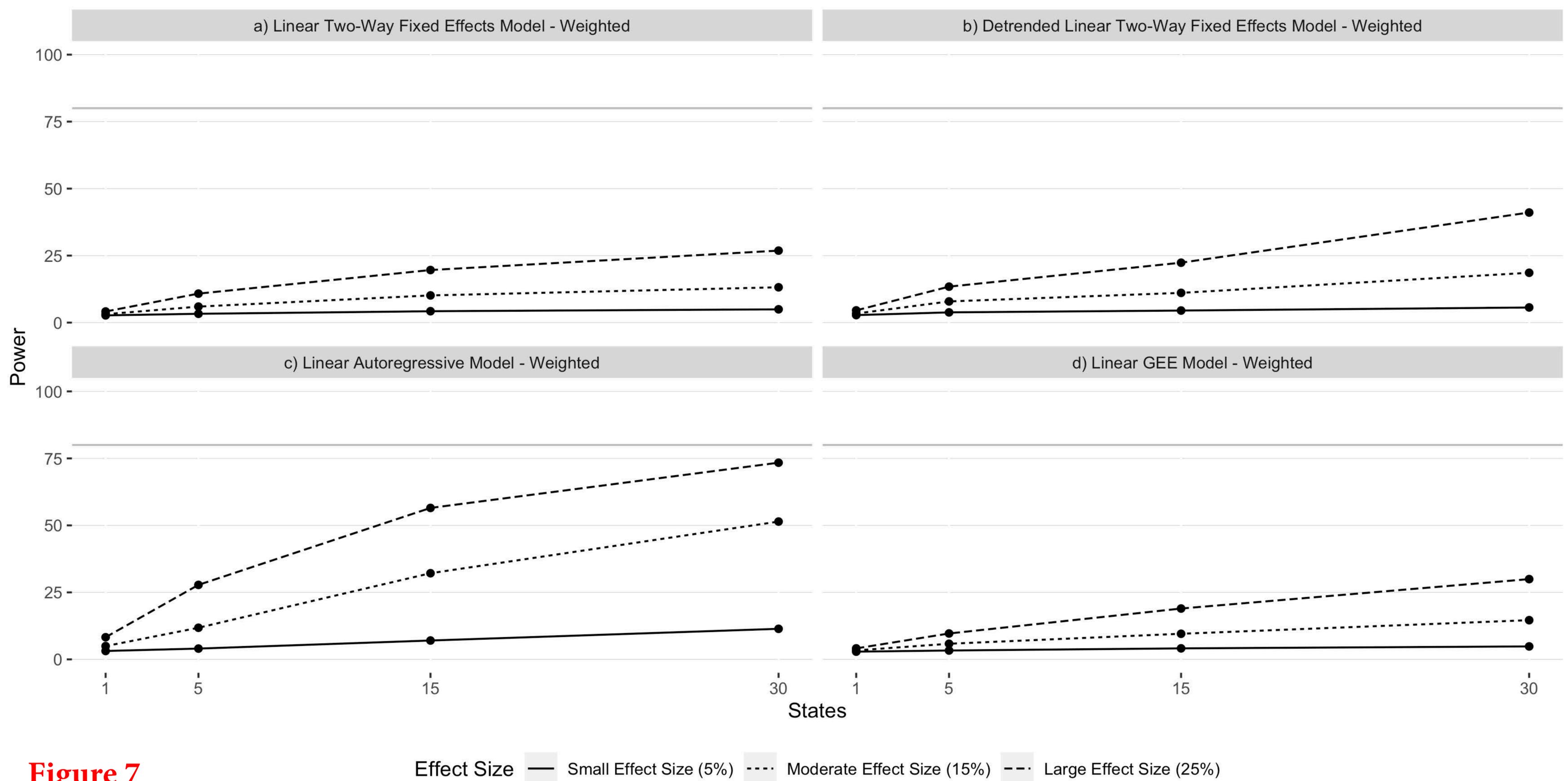

**Figure 7**

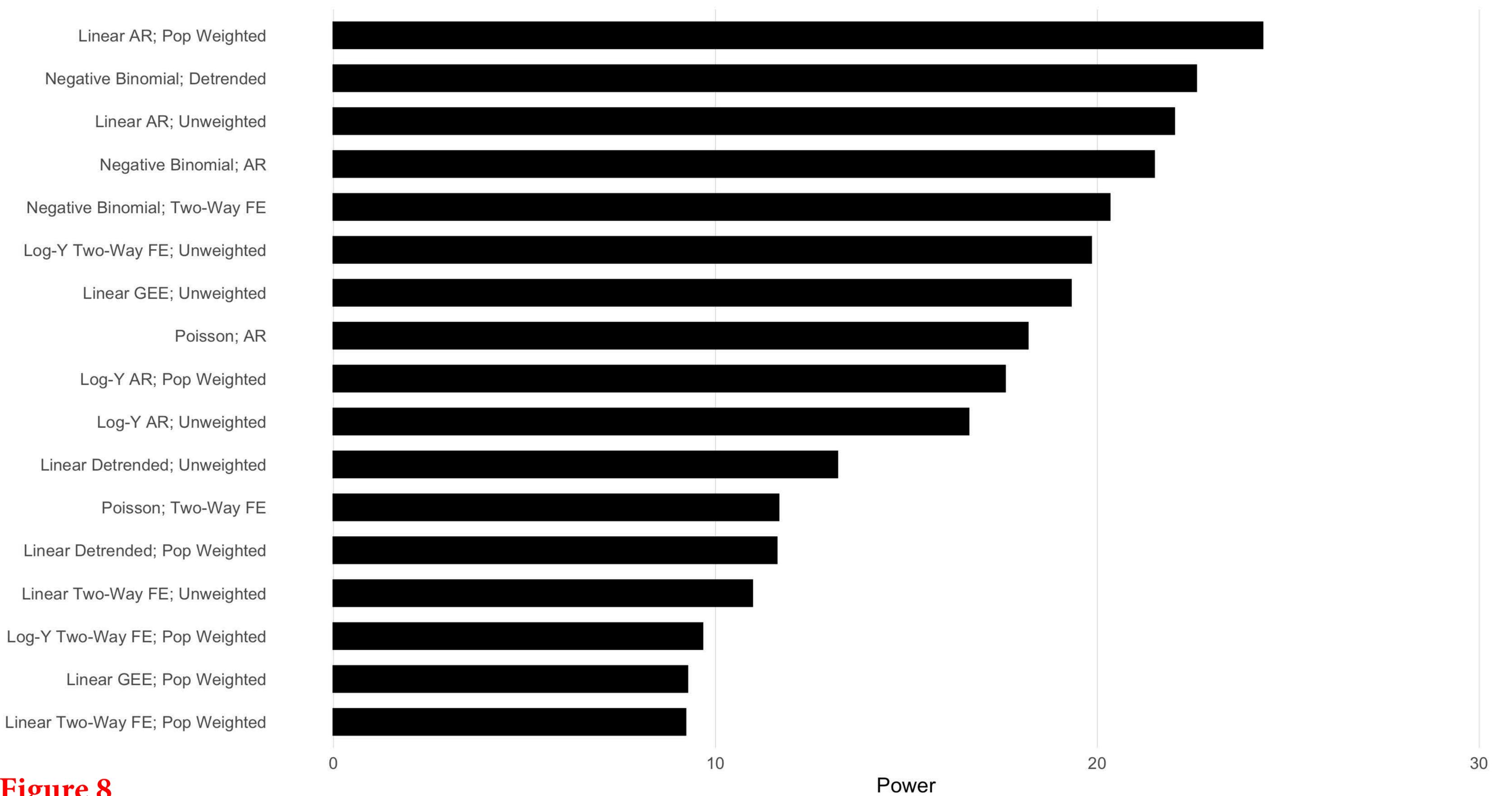

**Figure 8**

**Appendix:** Additional technical details and code

Standard error (SE) estimation: As noted, we estimated the SE in the three ways for each model (except GEE): (1) no adjustment; (2) Huber adjustment (robust estimators that attempt to adjust the SE for violations of distributional assumptions (White 1980; Zeileis 2004)); and (3) cluster adjustment ( adjustments to account for possible violations of the assumed independence of observations within states (White 1980; Zeileis 2004, 2006)). To illustrate the differences, we can write each adjustment in the following way (STATA Undated):

(1) No adjustment: This is just the normal ordinary least squares (OLS) estimator. The generic formula for a regression model with covariate matrix $\boldsymbol{X}$ capturing all the needed data information from the sample of observations can be written:

$$V_{OLS} = s^2 * (\boldsymbol{X'X})^{-1}$$

where $s^2 = (\frac{1}{N-k})\sum_{i=1}^{N} e_i^2$, $k$ denotes the number of parameters in a model and $e_i$ denotes the estimated residual for the i-th observation.

(2) Huber adjustment: This is the typical robust (unclustered) variance estimator and can be written as:

$$V_{huber} = s^2 * (\boldsymbol{X'X})^{-1} * [\sum_{i=1}^{N} (e_i * x_i)' * (e_i * x_i)] * (\boldsymbol{X'X})^{-1}$$

where $x_i$ denotes the i-th observation's vector of covariate values going into the covariate matrix $\boldsymbol{X}$.

(3) Cluster adjustment: This can be written as:

$$V_{cluster} = (\boldsymbol{X'X})^{-1} * [\sum_{j=1}^{n_c} (u_j' * u_j)] * (\boldsymbol{X'X})^{-1}$$

where $u_j = \sum_{j \, cluster} (e_i * x_i)$ and $n_c$ denotes the number of clusters.

Our simulations were run in R so we used the vcovHC package (RDocumentation Undated) to estimate Huber adjusted SEs by running the following command on the model output (here denoted by m1): vcovHC(m1, type="HC0"). As noted in R, type = "HC0" corresponds to the traditional Huber-White robust SE estimator. We confirmed that this Huber SE adjustment is the same as the "robust" SE option used by Stata (vce (robust)). To estimate the cluster adjusted SE, we utilized the following code to obtain the needed sandwich adjusted SE:

```
robust.se <- function(model, cluster){
  require(sandwich)
  require(lmtest)
  M <- length(unique(cluster))
  N <- length(cluster)
  K <- bigel$rank
  dfc <- (M/(M - 1)) * ((N - 1)/(N - K))
  uj <- apply(estfun(bigel), 2, function(x) tapply(x, cluster, sum));
  rcse.cov <- dfc * sandwich(bigel, meat = crossprod(uj)/N)
  rcse.se <- coeftest(bigel, rcse.cov)
  return(list(rcse.cov, rcse.se))}

clustervar<-mapply(paste,"State.",x$STATE,sep="")
m1$coefficients<-robust.se(m1,clustervar)[[2]]
```

We confirmed that the results from this function produces identical SE estimates as the cluster adjustment command used in Stata (vce (cluster state)).

Weighting: State-population weights were treated as analytic weights, not survey weights, within the analyses.

Model differences: Note that since the AR models use lagged outcomes in the regression model, they utilized one less year of data from the time series than the other models considered.

Code. We are currently finishing a codebase that will be placed on GitHub allowing for easy replication and extension of our code for running the simulations in this study. Here, we copy a sample run for one model (the unweighted linear two-way fixed effects model) over the null and assuming a 5% effect of the simulated law.

```
rm()
library(DataCombine)
library(MASS)
library(sandwich)
library(lmtest)
############################################
##Functions
############################################
#function needed for slow coding
const<-function(m)
{
  v=0
  if(m!=0)
  {
    for(i in 1:m)
    {
      v=v+i
    }
  }
  return(v)
}

#function needed for slow coding
slow.acting<-function(month,length,monthly.effect)
{
  top=length-1
  total.times<-c(1:top) #creating length+1 spline values for the slow acting time span
```

```r
  #compute year 1 average effect

  Fraction.year.enacted<-(13-month)/12

  Average.effect.while.enacted =0.5*Fraction.year.enacted*(1/length)

  Average.effect.over.year1 = Fraction.year.enacted*Average.effect.while.enacted

  values.midyrs<-Average.effect.over.year1 + total.times*(1/length)

  value.last.yr<-((13-month)*1+(month-1)-const(month-1)*monthly.effect)/12

  values=c(Average.effect.over.year1,values.midyrs,value.last.yr)

  return(values)

}

# calculate mean squared error

mse<-function(x)

{

  return(mean(x^2,na.rm=T))

}

#Set p-values so denote if result statistically signifcant at alpha = 0.05 level

#0 for p>=0.05 and 1 for p<0.05

pval.bin<-function(p)

{

  p[p<0.05]=1

  p[p!=1]=0

  return(p)

}

# Calculate correction factor for standard error

corr.factor<-function(t.stats)

{

  f.stats=(t.stats)^2

  f.stats=sort(f.stats)

  high.cut=0.95*iters

  femp95=f.stats[high.cut]

  freal=qf(.95,1,Inf)

  corr.factor=sqrt(femp95/freal)

  return(corr.factor)

}
```

```r
#formula for correcting p-values using correction factor

adj.ps<-function(regn.coeffs,ses,cf)

{

  adj.ses=sqrt(ses)*cf

  low95=regn.coeffs-1.96*adj.ses

  high95=regn.coeffs+1.96*adj.ses

  new.p=rep(0,iters)

  for(i in 1:iters)

  {

   if(low95[i]<0&high95[i]>0)

   {

    new.p[i]=1

   }else{

    new.p[i]=0

   }

  }

  return(new.p)

}

#type S

type.s<-function(betas,pvals,effect.direction)

{

  if(length(betas[pvals<0.05])!=0)

  {

   if(effect.direction=="neg"){

    a=length(betas[betas>0&pvals<0.05])/length(betas[pvals<0.05])

   }else{

    a=length(betas[betas<0&pvals<0.05])/length(betas[pvals<0.05])

   }

  }else{

   a=0

  }

  return(a)

}
```

```
#correct rejection rate

test.cf<-function(regn.coeffs,ses,cf,effect.direction)

{

  adj.ses=sqrt(ses)*cf

  low95=regn.coeffs-1.96*adj.ses

  high95=regn.coeffs+1.96*adj.ses

  new.p=rep(0,iters)

  for(i in 1:iters)

  {

   if(low95[i]<0&high95[i]>0)

    {

     new.p[i]=0

    }else{

     new.p[i]=1

    }

  }

  #switch findings in the incorrect direction to 0's

  if (effect.direction == "pos"){

    new.p[new.p==1®n.coeffs<0]=0

   }else{

    new.p[new.p==1®n.coeffs>0]=0

   }

  return(sum(new.p)/iters) #should be ~0.05

}

#needed for performing cluster adjustment to standard errors

robust.se <- function(model, cluster){

  require(sandwich)

  require(lmtest)

  M <- length(unique(cluster))

  N <- length(cluster)

  K <- model$rank

  dfc <- (M/(M - 1)) * ((N - 1)/(N - K))

  uj <- apply(estfun(model), 2, function(x) tapply(x, cluster, sum));
```

```r
  rcse.cov <- dfc * sandwich(model, meat = crossprod(uj)/N)

  rcse.se <- coeftest(model, rcse.cov)

  return(list(rcse.cov, rcse.se))

}

###############################

#Simulation function

###############################

# The main simulation generator and output function

run.sim = function(code.speed, effect.direction){

  #creating matrix to hold 4 key regression results a

  #Column 1 = estimated effect (regression coefficient)

  #Column 2 = estimated variance

  #Column 3 = t-statistic

  #Column 4 = p-value

  stats.matrix1=list(matrix(0,iters,4),matrix(0,iters,4),matrix(0,iters,4),matrix(0,iters,4))

  stats.matrix1h=list(matrix(0,iters,4),matrix(0,iters,4),matrix(0,iters,4),matrix(0,iters,4))

  stats.matrix1cl=list(matrix(0,iters,4),matrix(0,iters,4),matrix(0,iters,4),matrix(0,iters,4))

  stats.matrix1hcl=list(matrix(0,iters,4),matrix(0,iters,4),matrix(0,iters,4),matrix(0,iters,4))

  #use same seed for all runs so run on same simulated dataset for each model

  set.seed(1234567)

  #outer loop covers 4 different treated states sample sizes

  for(j in 1:4)

  {

    n.trt=n.states[j]

    #inner loop created the needed iters of simulated datasets

    for(k in 1:iters)

    {

      #Create vector of state names to sample from

      state.names=as.character(unique(x$STATE))

      #randomly sample the exposed/treated states

      z=sample(state.names,n.trt,replace=FALSE)

      #randomly sample the year the law was enacted for each treated states
```

```r
years.enacted=sample(c(2002:2013),n.trt,replace=TRUE)

#randomly sample the month the law was enacted for each treated state

month.enacted=sample(c(1:12),n.trt,replace=TRUE)

#create levels coding

x$levels.coding=rep(0,nrow(x))

if (code.speed == "slow") {

  #Slow coding - assumes it takes 3 years for a law to become fully effective

  length=3

  #loops through for each treated/exposed state to create the needed slow coding levels coding

  for(s in 1:n.trt)

  {

    month=month.enacted[s]

    values=slow.acting(month,length,monthly.effect = (1/length)/12)

    mark=length+1

    mark2=years.enacted[s]+length

    check=2016-years.enacted[s]

    if(check>=length(values))

    {

      x$levels.coding[x$STATE==z[s]&x$YEAR>=years.enacted[s]][1:mark]=values

      x$levels.coding[x$STATE==z[s]&x$YEAR>mark2]=1

    }else{

      hold=check+1

      x$levels.coding[x$STATE==z[s]&x$YEAR>=years.enacted[s]][1:hold]=values[1:hold]

    }

  }

  #Creating change levels coding for models for treated/exposed states

  x$ch.levels.coding=rep(0,nrow(x))

  for(s in 1:n.trt)

  {

    levels=x$levels.coding[x$STATE==z[s]]

    levels.shifted=c(0,levels[-length(levels)])

    x$ch.levels.coding[x$STATE==z[s]]=levels-levels.shifted

  ]
```

```r
    }else{
    if (code.speed == "instant"){
      #Instantaneous coding version
      for(s in 1:n.trt)
      {
        x$levels.coding[x$STATE==z[s]&x$YEAR==years.enacted[s]]=(12-month.enacted[s]+1)/12
        x$levels.coding[x$STATE==z[s]&x$YEAR>years.enacted[s]]=1
      }

      #Creating change levels coding
      x$ch.levels.coding=rep(0,nrow(x))
      for(s in 1:n.trt)
      {
        levels=x$levels.coding[x$STATE==z[s]]
        levels.shifted=c(0,levels[-length(levels)])
        x$ch.levels.coding[x$STATE==z[s]]=levels-levels.shifted
      }
    }
  }

#GENERATING OUTCOMES
  if(effect.direction !="null"){
    ##################
    #Introduce treatment effects to state observations
    ##################
    if (link == "linear"){
    x$cr.adj=x$Crude.Rate+te*x$levels.coding
    }
    if (link == "log-lin"){
    x$logY.adj=log(x$Crude.Rate+x$Crude.Rate*(te-1)*x$levels.coding)
    x$cr.adj=exp(x$logY.adj)
    }
    if (link == "log"){
    x$deaths.adj=x$Deaths+x$Deaths*(te-1)*x$levels.coding
```

```r
    x$deaths.adj=round(x$deaths.adj)

    x$cr.adj=(x$deaths.adj*100000)/x$POPULATION

    }

#need lags to be computed on new adjusted crude rates as potential control covariate in models

mark1=dim(x)[2]+1

x <- slide(x, Var = "cr.adj", GroupVar = "STATE", slideBy = -1)

colnames(x)[mark1] <- "lag1"

#x$lag1 = x$cr.adj.lag1

    }else{

        x$cr.adj=x$Crude.Rate

        x$deaths.adj=x$Deaths

    x$lag1 = x$crude.rate.lag1

        x$logY.adj=log(x$Crude.Rate)

        x$lag1 = x$cr.adj.lag1

    }

######################################################

# Insert Regression Model Here - Illustrative Example

# the formula line below can be changed to include different effects,

# including lags (use variable "lag1"), change-levels coded effects variables.

# the model line can substitute other models.

  #let's test two way fixed effects WITHOUT population weights

m1=lm(cr.adj~levels.coding+as.factor(YEAR)+as.factor(STATE)+ UNEMPLOYMENTRATE,data=x)

######################################################

    #store results

    stats.matrix1[[j]][k,1]= summary(m1)$coefficients[2,1] #regression coefficient

    stats.matrix1[[j]][k,2]= summary(m1)$coefficients[2,2]^2 #se^2

    stats.matrix1[[j]][k,3]= summary(m1)$coefficients[2,3] #t-statistics

    stats.matrix1[[j]][k,4]= summary(m1)$coefficients[2,4] #p-value

#Coding for implementing adjustments to standard errors

    #Huber adjustment requires library("sandwich")

    cov.m1<- vcovHC(m1, type="HC0")

    std.err <- sqrt(diag(cov.m1))

    stats.matrix1h[[j]][k,1]= coef(m1)[2]
```

```
stats.matrix1h[[j]][k,2]= std.err[2]^2 #var

stats.matrix1h[[j]][k,3]= coef(m1)[2]/std.err[2]

stats.matrix1h[[j]][k,4] =2*pnorm(abs(coef(m1)/std.err), lower.tail=FALSE)[2]

# Arellano

cov.m1<- vcovHC(m1, type="HC1",cluster="STATE",method="arellano")

std.err <- sqrt(diag(cov.m1))

stats.matrix1hcl[[j]][k,1]= coef(m1)[2]

stats.matrix1hcl[[j]][k,2]= std.err[2]^2 #var

stats.matrix1hcl[[j]][k,3]= coef(m1)[2]/std.err[2]

stats.matrix1hcl[[j]][k,4] =2*pnorm(abs(coef(m1)/std.err), lower.tail=FALSE)[2]

#Cluster adjustment only

#Create the new variable with appropriate level names.

clustervar<-mapply(paste,"State.",x$STATE,sep="")

#Save the coefficient test output to an element in the model object

m1$coefficients<-robust.se(m1,clustervar)[[2]]

stats.matrix1cl[[j]][k,1]= m1$coefficients[2,1] #bias

stats.matrix1cl[[j]][k,2]= m1$coefficients[2,2]^2 #var

stats.matrix1cl[[j]][k,3]= m1$coefficients[2,3]

stats.matrix1cl[[j]][k,4]= m1$coefficients[2,4] #p-value

#####################################################

# remove generic lag variable

x = x[, -which(names(x) == "lag1")]

print(k)

} #ends k loop

print(j)

} #ends j loop

#######################################################

#Compute Summary Statistics for runs

#######################################################

if (effect.direction == "null"){

#expanding so this holds 16 rows for 4 n.trt times 4 SE models

stats1=matrix(0,16,5)

#loop through 4 sample sizes for the number of treated states
```

```r
    cols=c(1,2,5)

    for(j in 1:4)

    {

mark1=(j-1)*4+1

mark2=mark1+1

mark3=mark2+1

mark4=mark3+1

    #Computes Type I Error

stats1[mark1,4]=mean(pval.bin(stats.matrix1[[j]][,4]))

stats1[mark2,4]=mean(pval.bin(stats.matrix1h[[j]][,4]))

stats1[mark3,4]=mean(pval.bin(stats.matrix1cl[[j]][,4]))

stats1[mark4,4]=mean(pval.bin(stats.matrix1hcl[[j]][,4]))

    #Computes Simple Mean Summaries for the other columns

stats1[mark1,cols]=apply(stats.matrix1[[j]][,1:3],2,mean)

stats1[mark2,cols]=apply(stats.matrix1h[[j]][,1:3],2,mean)

stats1[mark3,cols]=apply(stats.matrix1cl[[j]][,1:3],2,mean)

stats1[mark4,cols]=apply(stats.matrix1hcl[[j]][,1:3],2,mean)

    #Computes MSE under null

stats1[mark1,3]=mse(stats.matrix1[[j]][,1])

stats1[mark2,3]=mse(stats.matrix1h[[j]][,1])

stats1[mark3,3]=mse(stats.matrix1cl[[j]][,1])

stats1[mark4,3]=mse(stats.matrix1hcl[[j]][,1])

    }

    file1=paste("Summaries_Null_",code.speed,"_", model.name,".csv",sep="")

    file1=paste("Summaries_Null_",code.speed,"_", model.name,".csv",sep="")

    n.states.exp=c(rep(n.states[1],4),rep(n.states[2],4,),rep(n.states[3],4),rep(n.states[4],4))

    se.adj=c(rep(c("none","Huber","Cluster","Huber-Cluster"),4))

    stats1=as.data.frame(cbind(n.states.exp,se.adj,stats1))

    names(stats1)<-c("n.trt","se.adj","RegnCoeff","AveModelSE","MSE","TypeI","Tstat")

    write.table(stats1,file=file1,sep=",",row.names=FALSE)

    #Compute Correction Factors

    stats1.cf=rep(0,16)

    #compute for each number of treated/exposed states
```

```r
   for(j in 1:4)

   {

mark1=(j-1)*4+1

mark2=mark1+1

mark3=mark2+1

mark4=mark3+1

stats1.cf[mark1]=corr.factor(stats.matrix1[[j]][,3])

stats1.cf[mark2]=corr.factor(stats.matrix1h[[j]][,3])

stats1.cf[mark3]=corr.factor(stats.matrix1cl[[j]][,3])

stats1.cf[mark4]=corr.factor(stats.matrix1hcl[[j]][,3])

   }

   file2=paste("Correction_Factors_",code.speed,"_",model.name,".csv",sep="")

   write.table(stats1.cf,file2,sep=",",row.names=F)

# if instead it is a positive or negative effect run...

 }else{

   stats1=matrix(0,16,3)

   #Calculate bias

   if (link=="linear"){

     for(j in 1:4)

     {

mark1=(j-1)*4+1

mark2=mark1+1

mark3=mark2+1

mark4=mark3+1

             #bias

             tot.pop=sum(as.numeric(x$POPULATION))

             ave.pop.per.yr=tot.pop/length(unique(x$YEAR))

             APS = ave.pop.per.yr/100000

             TE = target.d

stats1[mark1,1]=mean(stats.matrix1[[j]][,1]*APS-TE)

stats1[mark2,1]=mean(stats.matrix1h[[j]][,1]*APS-TE)

stats1[mark3,1]=mean(stats.matrix1cl[[j]][,1]*APS-TE)
```

```
stats1[mark4,1]=mean(stats.matrix1hcl[[j]][,1]*APS-TE)

    }

   }else{

    for(j in 1:4)

    {

mark1=(j-1)*4+1

mark2=mark1+1

mark3=mark2+1

mark4=mark3+1

           tot.deaths=sum(x$Deaths)

           ave.per.yr=tot.deaths/length(unique(x$YEAR))

       ADPY = ave.per.yr

           TE = target.d

stats1[mark1,1]=mean((exp(stats.matrix1[[j]][,1])-1)*ADPY-TE)

stats1[mark2,1]=mean((exp(stats.matrix1h[[j]][,1])-1)*ADPY-TE)

stats1[mark3,1]=mean((exp(stats.matrix1cl[[j]][,1])-1)*ADPY-TE)

stats1[mark4,1]=mean((exp(stats.matrix1hcl[[j]][,1])-1)*ADPY-TE)

    }

   }

   ########################

   #adjusted power & adjusted type S error - requires correction Factor

   file2=paste("Correction_Factors_",code.speed,"_",model.name,".csv",sep="")

   cfs=read.table(file2,sep=",",h=T)

   for(j in 1:4)

   {

mark1=(j-1)*4+1

mark2=mark1+1

mark3=mark2+1

mark4=mark3+1

    #power

stats1[mark1,2]=test.cf(stats.matrix1[[j]][,1],stats.matrix1[[j]][,2],cfs$x[mark1],effect.direction)

stats1[mark2,2]=test.cf(stats.matrix1h[[j]][,1],stats.matrix1h[[j]][,2],cfs$x[mark2],effect.direction)

stats1[mark3,2]=test.cf(stats.matrix1cl[[j]][,1],stats.matrix1cl[[j]][,2],cfs$x[mark3],effect.direction)
```

```r
stats1[mark4,2]=test.cf(stats.matrix1hcl[[j]][,1],stats.matrix1hcl[[j]][,2],cfs$x[mark4],effect.direction)

  #type S error

stats1[mark1,3]=type.s(stats.matrix1[[j]][,1],adj.ps(stats.matrix1[[j]][,1],stats.matrix1[[j]][,2],cfs$x[mark1]),effect.direction)

stats1[mark2,3]=type.s(stats.matrix1h[[j]][,1],adj.ps(stats.matrix1h[[j]][,1],stats.matrix1h[[j]][,2],cfs$x[mark2]),effect.direction)

stats1[mark3,3]=type.s(stats.matrix1cl[[j]][,1],adj.ps(stats.matrix1cl[[j]][,1],stats.matrix1cl[[j]][,2],cfs$x[mark3]),effect.direction)

stats1[mark4,3]=type.s(stats.matrix1hcl[[j]][,1],adj.ps(stats.matrix1hcl[[j]][,1],stats.matrix1hcl[[j]][,2],cfs$x[mark4]),effect.direction)

    }

    if (link=="linear"){

      ave.coefficient=stats1[,1]+TE

      bt=abs(te)

    }else{

      ave.coefficient=stats1[,1]+TE

      bt=abs(log.te)

    }

    ll = list(stats1,ave.coefficient,bt)

    file3=paste("Results_",effect.direction,"_",code.speed,"_", model.name,".Rdata",sep="")

    save(ll,file=file3)

  }

}

###########################################

  #Step 1: Prepare the data for the simulation

  ###########################################

  setwd("FILL IN")

  load("optic_sim_data_exp.Rdata")

  ##############################################################################

  #Step 2. Set general simulation parameters

  ##############################################################################

  # number of iterations

  iters = 5000

  # effect coding, slow or instant

  code.speed = c("instant","slow")[1]   #select coding scheme

  # link type

  # select what type of effect modeling - linear = 1; log-linear = 2; log/count = 3
```

```r
link = c("linear", "log-lin", "log")[1]

# name for current model

model.name = "Opioid_Mortality_Runs_linear_2wayfe_unwt_smES"

#Creating 4 variations in sample size that we study

#Number of exposed/treated states = 1, 5, 15, and then 30

n.states=c(1,5,15,30)

##########################################################

# Step 3. Run null models (instant and slow), and positive

# and negative models (instant and slow)

##########################################################

# cycle through simulations of the null, and positive and negative effects

for (i in c("null","pos","neg")){

  if (i != "null"){

    #Generate effect magnitudes

    if (link=="linear"){

      #SIMULATING NONZERO POSITIVE EFFECTS FOR LINEAR MODELS

      #first we figured out what % change equals ~700 deaths

      tot.pop=sum(as.numeric(x$POPULATION))

      ave.pop.per.yr=tot.pop/length(unique(x$YEAR))

      APS = ave.pop.per.yr/100000

      target.d=700

      TE = target.d

      te=TE/APS

      if (i=="neg")

              {

              te=-te

              target.d=-target.d

              }

    }else{

      #SIMULATING NONZERO EFFECTS FOR COUNT MODELS AND LOG(Y) MODELS

      #first we figured out what % change equals ~700 deaths

      tot.deaths=sum(x$Deaths)

      ave.per.yr=tot.deaths/length(unique(x$YEAR))
```

```
    target.d=700

    percent.change=target.d/ave.per.yr

    if (i=="neg"){

     delta=1-percent.change

              target.d=-target.d

    }else{

     delta=1+percent.change

    }

    te=delta

    log.te=log(delta)

  }

 }

 # for each null, positive, or negative effect

 # cycle through simulations with instant and slow coding

 for (j in c("instant","slow")){

   dummy = run.sim(effect.direction=i, code.speed = j)

 }

}

############################################################

# Step 4. Organize resulting data

############################################################

for (i in c("instant","slow")){

 file3=paste("Results_","neg","_",i,"_", model.name,".Rdata",sep="")

 load(paste(file3,sep=""))

 ave.coefficient.neg = ll[[2]]

 results.neg.bias  = ll[[1]][,1]

 results.neg.power = ll[[1]][,2]

 results.neg.typeS = ll[[1]][,3]

 file3=paste("Results_","pos","_",i,"_", model.name,".Rdata",sep="")

 load(paste(file3,sep=""))

 ave.coefficient.pos = ll[[2]]
```

```r
results.pos.bias  = ll[[1]][,1]

results.pos.power = ll[[1]][,2]

results.pos.typeS = ll[[1]][,3]

  if(link=="log"){bt.count=ll[[3]]} else{bt.linear=ll[[3]]}

if(link=="log-lin"){bt.count=ll[[3]]} else{bt.linear=ll[[3]]}

  #power

results.power=(results.neg.power+results.pos.power)/2

#type S

results.typeS=(results.neg.typeS+results.pos.typeS)/2

#bias

results.bias=(results.neg.bias+results.pos.bias)/2

results.magbias=(results.pos.bias-results.neg.bias)/2

n.states.exp=c(rep(n.states[1],4),rep(n.states[2],4,),rep(n.states[3],4),rep(n.states[4],4))

se.adj=c(rep(c("none","Huber","Cluster","Huber-Cluster"),4))

all.results=cbind(n.states.exp,se.adj,results.bias,results.magbias,results.typeS,results.power)

all.results<-as.data.frame(all.results)

names(all.results)<-c("n.states.exp","se.adj","results.bias","results.magbias","results.typeS","results.power")

file4 = paste("Results_NonZeroEffect_",i,"_",model.name,".csv",sep="")

write.table(all.results,file=file4,sep=",",row.names=FALSE)

}

#compile into 2 columns

file4 = paste("Results_NonZeroEffect_","slow","_",model.name,".csv",sep="")

slow.results = read.table(file4,sep=",",header=TRUE)

file4 = paste("Results_NonZeroEffect_","instant","_",model.name,".csv",sep="")

instant.results =read.table(file4,sep=",",header=TRUE)

results.power=cbind(instant.results$results.power,slow.results$results.power)

results.typeS=cbind(instant.results$results.typeS,slow.results$results.typeS)

results.bias=cbind(instant.results$results.bias,slow.results$results.bias)

results.magbias=cbind(instant.results$results.magbias,slow.results$results.magbias)

n.states.exp=c(rep(n.states[1],4),rep(n.states[2],4,),rep(n.states[3],4),rep(n.states[4],4))

  se.adj=c(rep(c("none","Huber","Cluster","Huber-Cluster"),4))

all.results=cbind(n.states.exp,se.adj,results.bias,results.magbias,results.typeS,results.power)

all.results<-as.data.frame(all.results)
```

```r
names(all.results) =c("n.states","se.adj","results.bias.instant","results.bias.slow","results.magbias.instant",
  "results.magbias.slow","results.typeS.instant","results.typeS.slow",
                      "results.power.instant","results.power.slow")
file4 = paste("All_Results_NonZeroEffect_",model.name,".csv",sep="")
write.table(all.results,file4,sep=",",row.names=F)
```